\documentclass[11pt]{article}
\usepackage{titlesec}
\titleformat{\paragraph}[runin]{\normalfont\itshape}{\theparagraph.}{.3em}{}[.]\titlespacing{\paragraph}{0pt}{1ex plus .1ex minus .2ex}{.5em}
\usepackage{amsmath,amssymb,bm}
\usepackage{mathabx}
\usepackage[T1]{fontenc}
\usepackage[utf8]{inputenc}
\usepackage{lmodern}
\usepackage{mathtools}
\usepackage{dsfont}
\pdfoutput=1
\usepackage[english]{babel}
\usepackage[letterpaper, hmargin=1in, top=1in, bottom=1.2in, footskip=0.6in]{geometry}
\usepackage{graphicx} % Allows including images
\usepackage{booktabs} % Allows the use of \toprule, \midrule and \bottomrule in tables
\usepackage{color}
\definecolor{aquamarine}{rgb}{0.5, 1.0, 0.83}
\definecolor{ao(english)}{rgb}{0.0, 0.5, 0.0}
\definecolor{armygreen}{rgb}{0.29, 0.33, 0.13}
\definecolor{awesome}{rgb}{1.0, 0.13, 0.32}
\definecolor{ballblue}{rgb}{0.13, 0.67, 0.8}
\definecolor{bittersweet}{rgb}{1.0, 0.44, 0.37}
\definecolor{blue}{rgb}{0.0, 0.0, 1.0}
\definecolor{brinkpink}{rgb}{0.98, 0.38, 0.5}
\definecolor{ballblue}{rgb}{0.13, 0.67, 0.8}
\definecolor{brightturquoise}{rgb}{0.03, 0.91, 0.87}
\definecolor{blue-green}{rgb}{0.0, 0.87, 0.87}
\definecolor{caribbeangreen}{rgb}{0.0, 0.8, 0.6}
\definecolor{cyan}{rgb}{0.0, 1.0, 1.0}
\definecolor{amber(sae/ece)}{rgb}{1.0, 0.49, 0.0}
\graphicspath{ {images/} }
\definecolor{vdarkred}{rgb}{0.6,0,0.2}

\usepackage[utf8]{inputenc}
\usepackage{lmodern}

\definecolor{vdarkred}{rgb}{0.6,0,0.2}
\definecolor{vdarkblue}{rgb}{0,0.2,0.6}
\usepackage[pdftex, colorlinks, linkcolor=vdarkblue,citecolor=vdarkred]{hyperref}

\author{	
J\"urg Fr\"ohlich\footnote{Email: juerg@phys.ethz.ch}\\
Institute for Theoretical Physics\\
ETH Zurich\\
8093 Zurich, Switzerland
}

\title{Gauge Invariance and Anomalies in Condensed Matter Physics}

\begin{document}

\maketitle

\begin{abstract}
\textit{This paper begins with a summary of a powerful formalism for the study of electronic states in
condensed matter physics called ``Gauge Theory of States/Phases of Matter.'' The chiral anomaly, 
which plays quite a prominent role in that formalism, is recalled. I then sketch an application of the 
chiral anomaly in $1+1$ dimensions to quantum wires. Subsequently, some elements of the quantum 
Hall effect in two-dimensional (2D) gapped (``incompressible'') electron liquids are reviewed. In particular, 
I discuss the role of anomalous chiral edge currents and of anomaly inflow in 2D gapped electron liquids
with explicitly or spontaneously broken time reversal, i.e., in Hall- and Chern insulators. The topological 
Chern-Simons action yielding the transport equations valid in the bulk of such systems and the associated 
anomalous edge action are derived. The results of a general classification of ``abelian'' Hall insulators are outlined. 
After some remarks on induced Chern-Simons actions, I sketch results on certain 2D chiral photonic wave guides. 
I then continue with an analysis of chiral edge spin-currents and the bulk response equations in time-reversal 
invariant 2D topological insulators of electron gases with spin-orbit interactions. The ``chiral magnetic effect'' 
in 3D systems and axion-electrodynamics are reviewed next. This prepares the ground for an outline of a 
general theory of 3D topological insulators, including ``axionic insulators''. Some remarks on Weyl 
semi-metals, which exhibit the chiral magnetic effect, and on Mott transitions in 3D systems with dynamical 
axion-like degrees of freedom conclude this review.}
\end{abstract}

\section{Gauge Theory of States of Matter}\label{Intro}
The purpose of this paper is to review some of the work on the subject of ``toplogical states of matter'' carried out 
by my collaborators and myself during the period from the late nineteen-eighties until the late nineties. 
My presentation is based on results published in \cite{F-Marchetti} -- \cite{Faddeev} and references given there. 
Our work has led to a general formalism, dubbed \textit{``Gauge Theory of States/Phases of Matter,''} 
that has turned out to be very useful in characterizing certain states of condensed matter, among them
\textit{``topological insulators.''} This formalism complements the \textit{Landau theory} of phases 

One of the insights at the root of our initial efforts was the realization that the \textit{chiral anomaly}, discovered by
particle theorists (see \cite{CAA}), can be used to predict the existence of chiral edge currents in two-dimensional
interacting electron gases exhibiting the quantum Hall effect, and that these edge currents can be described using 
results from the theory of current (Kac-Moody) algebras; see \cite{F-Kerler}. For the integer Hall effect observed in
two-dimensional systems of non-interacting electrons in an external magnetic field, the existence of
chiral edge currents had earlier been predicted by B.~Halperin in a celebrated paper \cite{Halperin}.\footnote{See also
\cite{FGW} for some mathematical aspects of the theory of edge currents in non-interacting electron liquids.}
Using the chiral anomaly in a general analysis of chiral edge currents in two-dimensional
electron gases exhibiting the quantum Hall effect, including interacting gases, was proposed by X.-G.~Wen 
in \cite{Wen} independently of our simultaneous efforts. 
Further comments placing ideas presented in the following into a wider
context, as well as plenty of references can be found in subsequent sections (see also \cite{Faddeev}). The subject covered
in this paper owes its advancement and its successes to the efforts of many colleagues. Some of
their work will be referred to or is quoted in papers referred to in the text that follows.\footnote{I offer my 
apologies to all those colleagues whose contributions may not be properly highlighted 
in my paper, due to my lack of knowledge of their work or oversights, or because they are too recent to
be included in this paper; (I am not familiar with very recent results on topological states of matter).
Results concerning topologically protected states of systems of \textit{non-interacting} 
quantum particles are neither reviewed, nor referred to in the following; but see \cite{ASS, topology}. They 
will surely be discussed extensively in other contributions to this volume.}

The analysis presented in this review is not meant to be mathematically rigorous, although the arguments 
described in the text or in the papers on which it is based are quite carefully constructed and, I think, convincing. 
It is hoped that the paper may help its readers to understand general features of the physics of 
topologically protected states (or phases) of condensed matter at very low temperatures. The following
picture is a metaphore for topologically non-trivial matter. 
\begin{center}
\textit{An example of topologically non-trivial matter: a tree with the topology of the torus}\\
\vspace{0.2cm}
\includegraphics[height=5.4cm]{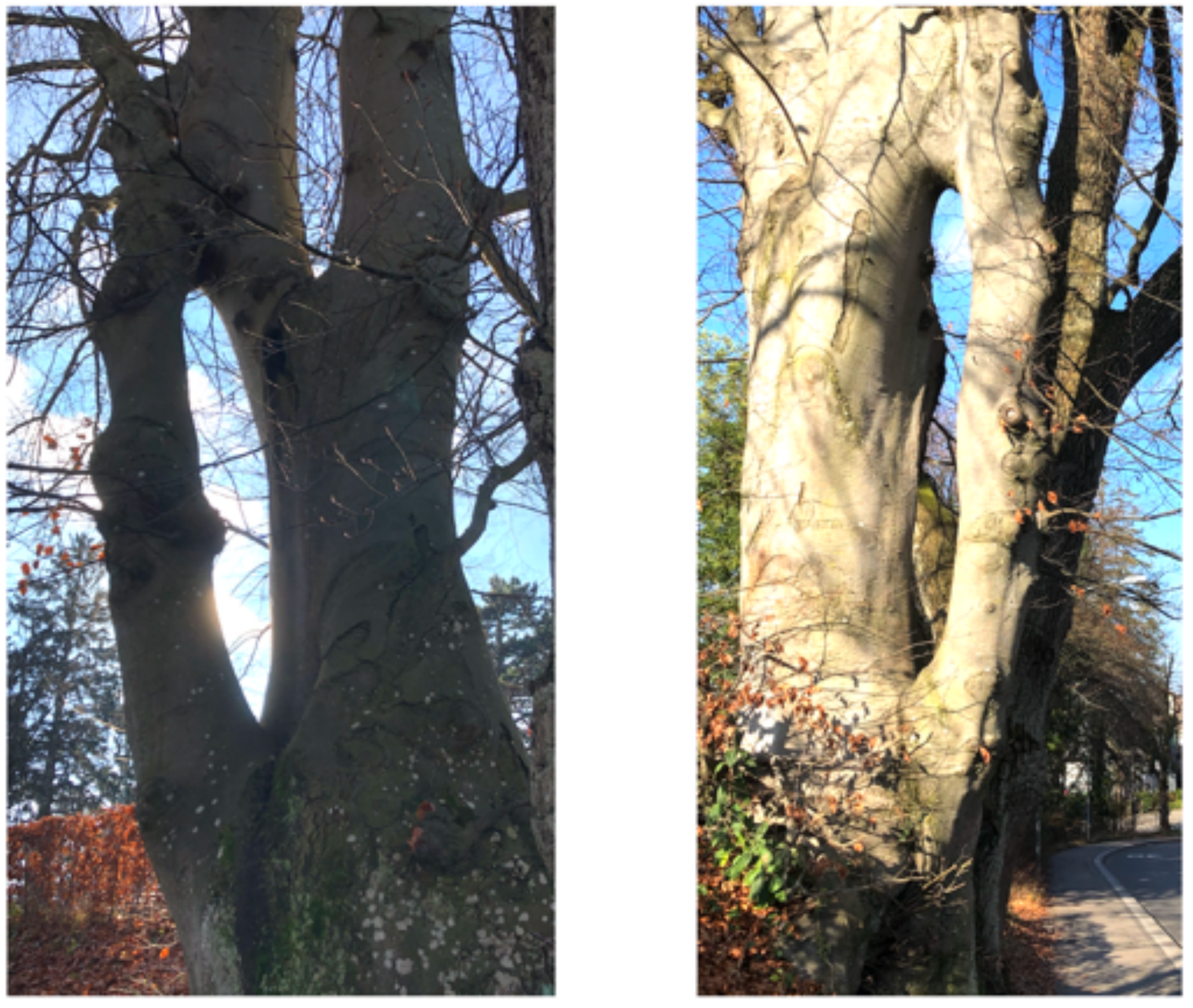}\\
{\tiny{Rigiblick -- Zurich, Switzerland}}
\end{center}
Here is a brief summary of the goals and of the contents of this paper. 
\begin{itemize}
\item{I will begin this review by describing concepts and results from gauge theory, current algebra and 
general relativity, with the purpose to outline the essence of the \textit{Gauge Theory of States/Phases 
of (Condensed) Matter.} This theory takes the place of \textit{Landau Theory} when the latter is not 
applicable, e.g., because there are no local order parameters available to characterize physical states 
of a system of condensed matter of interest. The \textit{Gauge Theory of States of Matter} reviewed in 
the following yields information on current Green functions, whence on transport coefficients (conductivities), 
which, thanks to \textit{Kubo formulae}, can be expressed as integrals over certain current Green functions

\textbf{Remarks.} (i) A system of condensed-matter physics with a \textit{``local order parameter''} 
is one whose phase diagram can be described with the help of an order-parameter field, $\varphi(x)$, 
where $x$ is a point in physical space. This field is most often a scalar or pseudo-scalar field with 
$N=1,2, \dots$ components, whose expectation value, $\langle \varphi(x) \rangle_{\xi}$, in a 
ground-state or equilibrium state $\big<(\cdot)\big>_{\xi}$ corresponding to a pure phase labelled by 
$\xi$ \textit{uniquely} identifies this phase, and whose correlators (Green functions) in such states enable one 
to study physically interesting fluctuations in the system at large distances and low energies. 
The systems studied in this paper do \textit{not} have local order parameters (in this sense of the expression). 
We will attempt to describe their ground-states and equilibrium states with the help of certain gauge field
theories. Special attention will be paid to \textit{``topological insulators''}, which -- it will turn out -- can be 
described by \textit{topological} gauge field theories, such as Chern-Simons theories and generalizations 
thereof. Some authors (see, e.g., \cite{Wen}) then speak of ``toplogical order,'' a somewhat fuzzy notion.\vspace{0.1cm}\\
(ii) I should mention that a ``change of perspective'' has recently been initiated in 
\cite{Gaiotto et al} and reviewed in \cite{McGreevy} (and references given there) based on 
developing the idea of ``higher-differential-form symmetry,'' which enables one to assign 
conserved charges to certain extended objects. With such generalizations of the concept of symmetry 
it appears to become possible to characterize most known equilibrium phases of matter by using an 
extended version of Landau theory derived from such generalized symmetries. -- I am not sufficiently 
knowledgeable to say much about the attempts alluded to here. But I tend to think that, when compared 
to the formalism described in this paper, the new perspective amounts mostly to a change of language.}
\item{I will review the following key ingredients of the \textit{Gauge Theory of States of Matter}:
\begin{enumerate}
\item{\textit{Effective Actions} $=$ generating functionals of connected current Green functions 
$\leftrightarrow$ formulae for transport coefficients, in particular \textit{conductivities}.}
\item{Implications of \textit{gauge invariance} ($\leftrightarrow$ current conservation, Ward identities), \textit{locality}, and 
\textit{power counting} (i.e., arranging different terms in an action functional according to their scaling dimensions) 
on the form of \textit{Effective Actions}.}
\item{\textit{Gauge anomalies} and their cancellations (anomaly inflow) $\leftrightarrow$ cooperation between
bulk- and edge/surface degrees of freedom $\leftrightarrow$ ``holography''.}
\end{enumerate}
}
\item{General features of the \textit{Gauge Theory of States of Matter} are described later in this section. 
I will then outline various applications of this theory to the characterization and classification of 
``topologically protected'' correlated \textit{bulk- and surface states} of \textit{interacting} systems of condensed 
matter, in particular of electron gases/liquids, which do not have local order parameters. Among such applications 
I will consider the following ones:\footnote{A fairly extensive list of references to relevant work will be given later in the text.}
\begin{enumerate}
\item{Conductance quantization in ideal quantum wires} 
\item{Theory of the Fractional Quantum Hall Effect}
\item{Theory of chiral states of light in certain planar wave guides}
\item{Time-reversal invariant planar ``topological'' insulators\\ 
and their chiral edge spin currents}
\end{enumerate}
}
\item{Subsequently I will proceed to review higher-dimensional cousins of the Quantum Hall Effect,\footnote{See 
\cite{F-Pedrini, BFR,BBFFRS}; they were later applied to studies in cold-atom physics -- see \cite{Zilber} and references given there.} 
3D topological insulators, and so-called \textit{Weyl semi-metals}, exhibiting the \textit{chiral 
magnetic effect}.\footnote{An effect discovered in a preliminary form in \cite{Vilenkin} and studied systematically in \cite{ACF}.} 
I will outline various consequences of this effect.}
\item{Applications of related ideas and methods in other areas of physics, in particular in cosmology, should also, yet will
not be discussed in this paper; but see \cite{ACF, F-Pedrini, BFR, BBFFRS}.}
\end{itemize}
In the remainder of this section, I introduce \textit{effective actions}, explain how their general form can be determined 
(under suitable assumptions) and elucidate their properties and uses. My discussion follows the one in \cite{Les Houches 94}.
 
We consider a quantum-mechanical system whose matter degrees of freedom are described by fields 
$\overline{\psi}, \psi, \dots$ over a space-time region, $\Lambda$, equipped with a metric, $g_{\mu\nu}$, of signature 
$(-1,1,1,1)$. The dynamics of the system is assumed to be derivable from an action functional 
\begin{equation}\label{action}
S(\overline{\psi}, \psi, ...; g_{\mu\nu}, A_{\mu})\,, 
\end{equation}
where $A=(A_{\mu})$ is the electromagnetic vector potential describing an external electromagnetic field (or 
some other gauge potential describing an external gauge field) acting on the system. We assume that there 
is a conserved vector current (density) $J^{\mu}$, with $\partial_{\mu}J^{\mu} =0$. If the current $J^{\mu}$ 
is charged, i.e., is carried by electrically charged degrees of freedom, then these degrees of freedom interact 
with the electromagnetic field. In the action functional \eqref{action} of the system, the charged fields are coupled to the electromagnetic field by ``minimal substitution,'' i.e., by replacing derivatives by covariant derivatives. 

The \textit{Effective Action} of the system at zero temperature on a space-time region $\Lambda$ with metric 
$g_{\mu\nu}$ in an external electromagnetic field with vector potential $A_{\mu}$ can then be constructed by, 
for example, functional integral methods.\footnote{Functional- or path integral methods were originally introduced by Dirac 
in \cite{Dirac}. An operator formalism to derive expressions for effective actions has been sketched elsewhere; 
see, e.g., \cite{F-Kerler, Les Houches 94}, and references given there.}
We consider the formal functional integral
\begin{align}\label{eff action}
S_{eff}(g_{\mu\nu}, A_{\mu}):= -i \hbar\, \text{ ln}\left( \int \mathcal{D}\overline{\psi} \mathcal{D} \psi \,exp[\frac{i}{\hbar}S(\overline{\psi}, \psi,...; g_{\mu\nu}, A_{\mu})]\right) +  \text{  (divergent) const.   } 
\end{align}
 A precise definition of the right side in \eqref{eff action} involves specifying initial and final field configurations; 
 e.g., ones corresponding to ground-states of the system (and renormalizing divergences when short-distance 
 regularizations are removed). Here I will not present any technical details concerning
 the precise definition of $S_{eff}$ and the choice of boundary conditions in \eqref{eff action}.\\

\noindent Next, I review some well known, important \textbf{properties of effective actions}.
 \begin{enumerate}
 \item{The variational derivatives of $S_{eff}$ with respect to $A_{\mu}$ are given by connected current Green functions:
 \begin{equation} \label{derivative-A}
 \frac{\delta S_{eff}(g_{\mu\nu}, A_{\mu})}{\delta A_{\mu}(x)}= \langle J^{\mu}(x) \rangle_{g, A}\,, 
 \end{equation}
 and
 \begin{equation}\label{derivative-g}
 \frac{\delta^{2}S_{eff}(g_{\mu\nu}, A_{\mu}) }{ \delta A_{\mu}(x)\, \delta A_{\nu}(y)} = \langle J^{\mu}(x)J^{\nu}(y) \rangle_{g,A}^{c}\,, \quad \text{ for }\,\,x \not= y\,,
  \end{equation}
 where $\langle (\cdot) \rangle_{g,A}$ denotes a time-ordered expectation in the presence of external fields 
 $g_{\mu \nu}$ and $A_{\mu}$. Formally this follows from Eq.~\eqref{eff action}.}
\item{We consider the effect of a gauge transformation, $ A_{\mu} \mapsto A_{\mu} + \partial_{\mu} \chi$, where 
$\chi$ is an arbitrary smooth function on $\Lambda$, on the effective action, $S_{eff}$. Using \eqref{derivative-A}
one finds that
\begin{equation}\label{gauge invariance}
\frac{\delta S_{eff}(g_{\mu\nu}, A_{\mu} + \partial_{\mu} \chi) }{\delta \chi(x)} = \partial_{\mu} \langle J^{\mu}(x) \rangle_{g,A}=0\
\end{equation}
vanishes, because $J^{\mu}$ is conserved. Thus, $S_{eff}$ is \textit{invariant under gauge transformations}.
 }
 \item{We may also vary $S_{eff}$ with respect to the metric $g_{\mu\nu}$:
 $$\frac{\delta S_{eff}(g_{\mu\nu}, A_{\mu})}{\delta g_{\mu\nu}(x)}= \langle T^{\mu\nu}(x) \rangle_{g,A}\,,$$
 where $T^{\mu\nu}$ is the energy-momentum tensor of the system. Using local energy-momentum conservation, 
 i.e., $\nabla_{\mu}T^{\mu\nu} =0$, we find that
$S_{eff}(g_{\mu\nu},A_{\mu})$ is invariant under space-time coordinate transformations.\\
A general (possibly curved) metric $g_{\mu\nu}$ can be used to describe certain defects (disclinations)
in a condensed-matter system.\\
Invariance of $S_{eff}$ under Weyl rescalings of the metric implies that 
$\langle T_{\mu}^{\mu}(x) \rangle_{g,A} \equiv 0 \leftrightarrow$ \textit{scale-invariance} (criticality) of the system.}
\item{In our analysis of ground-state properties of insulators it will be assumed that the external fields 
$g_{\mu\nu}$ and $A_\mu$ vary in space and time very slowly, so that an adiabatic theorem can be 
invoked in the description of the time-dependence of the state of the system under consideration, 
and that the Hamiltonians exhibit a spectral (or mobility) gap above the ground-state energy at all times. 
One may then argue that the zero-temperature connected current Green functions of such a system have 
good decay properties in space and time. In the \textit{scaling limit,} i.e., in the limit of very large distances and 
very low frequencies its effective action can then be expected to approach a functional that is a space-time integral 
of \textit{local, gauge-invariant polynomials} in $A_{\mu}$ and derivatives of $A_{\mu}$, as well as curvature 
quantities only depending on the metric $g_{\mu \nu}$.
These terms can be organized according to their scaling dimensions (power counting -- I recall that a gauge 
potential and a space-time derivative scale like an inverse length). Among these terms there may be 
\textit{topological} (metric-independent) action functionals that can be indicative of ``topological protection'' 
(or ``topological order'') of the states in question.}
\item{\textit{A generalization:} It is sometimes useful to consider more general effective actions involving 
\textit{non-abelian gauge fields} and to analyze the associated currents, which are only covariantly conserved. 
Non-abelian gauge fields may represent physical external fields; but they may also represent ``virtual'' or ``emergent''
ones merely serving to develop the response theory needed to determine transport coefficients, such as conductivities, 
associated with the currents corresponding to those gauge fields.\\
As an example of the relevance of non-abelian gauge fields in the study of condensed matter physics, I recall
that, in non-relativistic quantum theory, electron spin couples to $SU(2)$-gauge fields describing Zeeman
interactions\footnote{caused, e.g., by an external magnetic field or by the vorticity of a velocity field that generates  
the motion of an ionic background}, spin-orbit interactions and exchange interactions described by a \textit{Weiss 
exchange field} (see \cite{AFFS}). The non-relativistic quantum theory of electron gases/liquids turns out to be perfectly 
$U(1)_{em} \times SU(2)_{spin}$-gauge invariant (see \cite{Anandan, FS-RMP, Les Houches 94}). This observation 
is not only conceptually interesting; it can be used to, for example, analyze states of \textit{time-reversal invariant planar topological 
insulators} (see \cite{Faddeev} and Section 6) and chiral spin liquids (see, e.g., \cite{Les Houches 94}).}
\end{enumerate}

It turns out that properties 1 through 5 enable us to determine the general form of (bulk-) effective actions, $S_{eff}$, of 
insulators in the scaling limit and hence to derive formulae for transport coefficients, in particular certain conductivities.

\textit{\underline{Example}.} We consider an insulator with broken parity ($P$) and time-reversal ($T$) symmetries confined 
to a region, $\Omega$, of a flat 2D surface (e.g., an interface between an insulator and a semi-conductor) imbedded in physical space. The space-time of this system is defined to
be $\Lambda= \Omega \times \mathbb{R}$. The effective action of the system 
can then be argued to tend to
 $$S_{eff}(A) = \frac{\sigma_{H}}{2} \int_{\Lambda} A\wedge dA + \frac{1}{2}\int_{\Lambda} d^{2}x\, dt\,[\underline{E}(x)\cdot \varepsilon \underline{E}(x) - \mu^{-1} B(x)^{2} ] + \cdots \,,$$
as the scaling limit is approached, where $\sigma_{H}$ is the Hall conductivity, $\varepsilon$ is the tensor of dielectric 
constants, and $\mu$ is the magnetic permeability. (In the first term on the right side, $A$ is viewed as a differential
1-form, $d$ denotes exterior differentiation of differential forms, and $\wedge$ is the exterior product of differential forms.)
We note that the first term on the right side, i.e., the Chern-Simons action, is \textit{not} gauge-invariant if the space-time
region $\Lambda$ has a non-empty boundary, $\partial \Lambda= \partial \Omega \times \mathbb{R}$. This deficiency is
cured by a boundary action. The cooperation of the Chern-Simons action and the boundary action in preserving
gauge invariance may be viewed as an example of \textit{``holography''} (see Section 4).
 
The formalism sketched here can be generalized to be applicable to equilibrium states at positive temperature. 
(For further details concerning the material reviewed here, see \cite{FS-RMP, Les Houches 94, Faddeev}.)

\section{The chiral anomaly} \label{anomaly}

An excellent general reference for this section is \cite{CAA}.
We consider a system composed of relativistic, massless charged fermions in a (flat) space-time region, $\Lambda$, of dimension $2n, n=1,2,\dots$. The (operator-valued, quantum-mechanical) vector current of this system is denoted by 
$J^{\mu}$ and the axial-vector current by $J^{\mu}_{5}$. The vector current is conserved, i.e.,
$$\partial_{\mu}J^{\mu}=0\,\, \leftrightarrow \,\text{  gauge invariance of theory}.$$
But the axial-vector current tends to be \textit{anomalous.} In two space-time dimensions, for massless fermions,
\begin{equation}\label{ano}
\partial_{\mu}J^{\mu}_{5}= \frac{\alpha}{2\pi} E, \quad \alpha:= \frac{e^{2}}{\hbar}, \quad [J^{0}_{5}(\vec{y},t), J^{0}(\vec{x},t)] 
= -i \,\frac{\alpha}{2\pi} \delta^{'}(\vec{x}-\vec{y}),
\end{equation}
where $\alpha$ is the finestructure constant, and $E$ is the external electric field.\footnote{These formulae can be traced back
to Tomonaga's work on 1D electron liquids.}\\
In four dimensions, for massless fermions,
\begin{equation}\label{7}
\partial_{\mu}J^{\mu}_{5} = \frac{\alpha}{4\pi^{2}}\varepsilon^{\nu\lambda\kappa\rho}F_{\nu \lambda}F_{\kappa \rho} = \frac{\alpha}{2\pi^{2}} \vec{E}\cdot \vec{B}\,,
\end{equation}
and
\begin{equation}\label{Hamiltonian ano}
\big[J^{0}_{5}(\vec{y},t), J^{0}(\vec{x},t)\big]= i\frac{\alpha}{4\pi}\big(\vec{B}(\vec{y},t)\cdot \vec{\nabla}_{\vec{y}}\big)\,\delta(\vec{x}-\vec{y})\,,
\end{equation}
where $\vec{E}$ is the electric field and $\vec{B}$ is the magnetic induction. For massive fermions, there are terms
proportional to fermion masses contributing to $\partial_{\mu}J^{\mu}_{5}$.
The term $\vec{E}\cdot \vec{B}$ on the right side of \eqref{7}, the ``instanton density,'' is dual to
$\frac{1}{2}F\wedge F$ (where $F$ is interpreted as a 2-form), and $F\wedge F$ is the exterior dervative of 
the Chern-Simons 3-form, which expresses the \textit{helicity} of the electromagnetic field.

For later purposes we also introduce the left- and right chiral currents
\begin{equation}\label{chiral currents}
J^{\mu}_{\ell}:= \frac{1}{2}\big[J^{\mu}+J^{\mu}_{5}\big], \quad J^{\mu}_r:= \frac{1}{2}\big[J^{\mu}-J^{\mu}_{5}\big]\,.
\end{equation}
The commutation relations in \eqref{ano} and \eqref{Hamiltonian ano} then imply that
\begin{align}\label{ano commutators}
\big[J^{0}_{\ell/r}(\vec{y},t), J^{0}(\vec{x},t)\big] =& \pm i \frac{\alpha}{4\pi} \delta^{'}(\vec{x}-\vec{y})\,, \hspace{2.6cm} \text{ in } d=2\,,
\nonumber\\
\big[J^{0}_{\ell/r}(\vec{y},t), J^{0}(\vec{x},t)\big] =& \pm i\frac{\alpha}{8\pi}\big(\vec{B}(\vec{y},t)\cdot \vec{\nabla}_{\vec{y}}\big)\,
\delta(\vec{x}-\vec{y})\,, \quad \text{ in } d=4\,.
\end{align}

\noindent \textbf{Remark.}
It is sometimes of interest to study systems of matter in a non-inertially moving background -- e.g., electrons
in a rotating metallic cylinder, a Bose gas in a rotating trap, or the Earth's atmosphere or oceans. Suppose the 
motion of the background is generated by a (possibly time-dependent) velocity field, $\vec{V}$, on physical space. 
Studying the matter currents of such a system, one finds that there is an anomalous axial vector current 
satisfying an inhomogeneous continuity equation analogous to \eqref{7} and exhibiting anomalous commutators 
analogous to \eqref{Hamiltonian ano}, but with $M\vec{V}$ playing the role of $e\vec{A}$, hence $M\,\vec{\omega}$ 
playing the role of $e\vec{B}$, where $M$ is a mass scale and $\vec{\omega}=\vec{\nabla}\times \vec{V}$ is the 
vorticity of the velocity field $\vec{V}$.\footnote{In classical physics, the role of $e\vec{E}$ is played by 
$M\dot{\vec{V}} + \frac{M}{2} \vec{\nabla}\vert \vec{V}\vert^{2}$.}
Effects observed in systems imbedded in a moving background have been described in \cite{FS-RMP, Les Houches 94}. 
Some of them will be mentioned in Sections 4 and 7.\\

As a service to the reader we derive the formulae in \eqref{ano}, (setting $\hbar=1$). We consider a system of
massless fermions on 2D Minkowski space, $\Lambda$.
Let $\iota$ be the 1-form dual to the vector current density $J^{\mu}$. Then
$$\partial_{\mu}J^{\mu}=0 \,\,\Leftrightarrow\,\, d\,\iota =0.$$
Poincar\'e's lemma then implies that
$$\iota= \frac{Q}{2\pi} d \varphi, \quad \text{where }\,\varphi \, \text{ is a scalar field,}$$
and $Q$ is a quantity with the dimension of charge. Thus
\begin{equation}\label{current potential}
J^{\mu}= \frac{Q}{2\pi} \varepsilon^{\mu \nu}\partial_{\nu} \varphi
\end{equation}
In two space-time dimensions (using the metric $g_{\mu \nu}$ to raise and lower indices), 
the axial-vector current of a system of massless Dirac fermions is given by
\begin{equation}\label{axialcurrent}
J^{\mu}_{5}=\varepsilon^{\mu \nu}J_{\nu} \overset{\eqref{current potential}}{=} \frac{Q}{2\pi} \partial^{\mu} \varphi,
\end{equation}
see \cite{Seiler}. 

It is a general fact that, in two dimensions, the Hodge dual of the one-form $\iota$ is a one-form 
dual to an axial-vector current (and conversely). The following considerations apply to any system 
in two space-time dimensions with a conserved vector current $J^{\mu}$ and an axial-vector current 
$J^{\mu}_5$ related to $J^{\mu}$ by Hodge duality. Then
\begin{equation}\label{ax-current}
\partial_{\mu}J^{\mu}_5 = \frac{Q}{2\pi} \Box\, \varphi\,.
\end{equation}
Unitarity of the quantum theory of the field $\varphi$ suggests that it satisfies a field equation of the form
$$\Box\, \varphi = U'(\varphi, \partial_{\mu}\varphi, E)\,,$$
where $U'$ is a the variational derivative with respect to $\varphi$ of some functional $U$ of $\varphi, \partial_{\mu}\varphi$ and the external electric field $E$ (with $U$ bounded below).

In the following I consider systems in two space-time dimensions with the property that 
$$U'(\varphi, \partial_{\mu}\varphi, E\equiv 0)=0.$$ 
The field $\varphi$ then describes non-interacting massless modes. It follows that $J^{\mu}_5$ is conserved 
as long as the external electric field vanishes. An example where this remark applies is a 
Tomonaga-Luttinger liquid in the approximation where the dispersion law of the quasi-particles 
is linearized at the Fermi points. -- To summarize these remarks, we have that
\begin{equation}\label{ff}
\partial_{\mu}J^{\mu}_{5}=0 \,\, \overset{\eqref{axialcurrent}, \eqref{ax-current}}{\Leftrightarrow}\, \,\Box\, \varphi =0,
\end{equation}
provided that $E=0$; i.e., $\varphi$ is a \textit{massless free field}, which can be described by a Lagrangian quantum field 
theory with an action functional given by
\begin{equation}\label{ffaction}
S(\varphi)= \frac{1}{4\pi} \int_{\Lambda} d^{2}x\,\sqrt{-g(x)}\, \partial_{\mu}\varphi(x)\, \partial^{\mu} \varphi(x), \quad \text{with  }\,
x \equiv (t, \underline{x}) \in \Lambda\,.
\end{equation}
Assuming for simplicity that $\Lambda$ is flat, the ``momentum,'' $\varpi$, canonically conjugate to $\varphi$ is
given by
$$\varpi(x)= \frac{\delta S(\varphi)}{\delta(\partial_{0}\varphi(x))}=\frac{1}{2\pi} \frac{\partial \varphi(x)}{\partial t}
  = -Q^{-1} J^{1}(x).$$
By Eq.~\eqref{axialcurrent}, 
 $$J^{0}_{5}= Q \varpi, \quad J^{1}_{5}=\frac{Q}{2\pi} \frac{\partial \varphi}{\partial \underline{x}}\,.$$
The usual equal-time canonical commutation relations on Fock space,
$$[\varphi(t, \underline{x}),\varpi(t, \underline{y})] = i \,\delta(\underline{x}-\underline{y}),$$
then imply an ``anomalous current commutator''
\begin{equation}\label{ACC}
[J^{0}(t, \underline{x}), J^{0}_{5}(t, \underline{y})]= i \frac{Q^{2}}{2\pi} \delta'(\underline{x}-\underline{y})\,,
\end{equation}
see \eqref{ano}, with $Q^{2}=\alpha$. It is sometimes convenient to consider the \textit{chiral currents}\,
$ J^{\mu}_{\ell/r}= \frac{1}{2}\big[J^{\mu} \pm J^{\mu}_{5}\big].$
They can be used to express \textit{chiral Fermi fields} in terms of the bosonic fields $\varphi$ and $\varpi$. 
We define fields (so-called ``vertex operators'') $\psi_{\ell/r}^{(q)} (x)$ by setting
\begin{align}\label{cff}
 \psi_{\ell/r}^{(q)} (x) =& :\text{exp}\big\{\pm i4\pi \frac{q}{Q} \int_{\underline{x}}^{\infty} 
 \iota_{\ell/r}(x^{0}, \underline{y})\big\}:\nonumber\\
 =& :\text{exp} i q \big[\pm \varphi(x) + 2\pi \int_{\underline{x}}^{\infty} \varpi(x^{0}, \underline{y}) d\underline{y}\big]: \,,
\end{align}
where the double colons indicate Wick ordering with respect to the free scalar field with a \textit{positive} mass; (changing
the mass used in the definition of Wick ordering just changes the fields $\psi_{\ell/r}^{(q)} (x)$ by a constant factor).
When applied to a state in the Hilbert space of the system, the fields $ \psi_{\ell/r}^{(q)}$ change the electric charge
of the state by an amount $\,Q\cdot q$. The quantum statistics of the fields $ \psi_{\ell/r}^{(q)} $ is determined by the phase 
factor $e^{\pm i \pi q^{2}}$ appearing in the Weyl relations of the vertex operator on the right side of \eqref{cff}. 
Thus, if $q=1$ then $\psi_{\ell/r}^{(q)}$ is a Fermi field.\\
I hasten to add that the fields creating or annihilating \textit{electrons} in an interacting Tomonaga-Luttinger 
liquid \cite{Tomonaga, Luttinger} are \textit{not} the ones given in \eqref{cff}; rather, they are composed of 
left-chiral and right-chiral operators in a way determined by the strength of two-body interactions in the electron liquid 
(and by the requirements that they satisfy Fermi-Dirac statistics and create or annihilate an electric charge 
given by the elementary electric charge $e$).

To take into account the presence of an external electric field, $E$, the action functional $S(\varphi)$ in \eqref{ffaction} 
must be replaced by
\begin{align}\label{totaction}
 S(\varphi;A):=&\frac{1}{4\pi} \int_{\Lambda} \partial_{\mu}\varphi \partial^{\mu} \varphi\,d^{2}x + \int_{\Lambda} J^{\mu}A_{\mu}\,d^{2}x \nonumber\\
 =& \frac{1}{4\pi} \int_{\Lambda} \lbrace \partial_{\mu}\varphi \partial^{\mu} \varphi + 2Q\varepsilon^{\mu \nu} \partial_{\nu} \varphi A_{\mu}\rbrace d^{2}x \nonumber \\
=& \frac{1}{4\pi} \int_{\Lambda} \lbrace \partial_{\mu}\varphi \partial^{\mu} \varphi + 2Q\,\varphi E \rbrace d^{2}x\,,
\end{align}
where $E(x)=\varepsilon^{\mu \nu}(\partial_{\mu}A_{\nu})(x)$. This formula can be derived from the theory 
of massless free Dirac fermions coupled to a vector potential $A$ by convergent perturbation theory in the 
term $\int_{\Lambda} J^{\mu}A_{\mu} d^{2}x$ (see, e.g., \cite{Seiler}). From expression \eqref{totaction} one
easily derives the field equation for $\varphi$, which reads $\Box\, \varphi(x)=Q\,E(x)\,.$ Hence
\begin{equation}\label{ca}
\partial_{\mu} J^{\mu}_{5}= \frac{Q^{2}}{2 \pi} E(x)\,,
\end{equation}
which is the chiral anomaly in two space-time dimensions (interpreting $Q^{2}$ as the feinstructure constant $\alpha$).

Remarks on the chiral anomaly in four space-time dimensions will follow in Section 7.

\section{Conduction quantization in quantum wires}
The material presented in this section is taken form \cite{ACFr} and \cite{ACF}. We consider a
physical system consisting of a very long wire that contains a 1D interacting electron gas 
described by a Tomonaga-Luttinger liquid (setting $Q=-e$ in \eqref{current potential}) connected 
to electron reservoirs on the left end and the right end of the wire.
We assume that there are no back-scattering processes converting left-moving quasi-particles (charge-density
fluctuations) into right-moving ones, or conversely. Very close to the Fermi-points of the electron gas, the 
electron dispersion law can be linearized, and quasi-particles can be described as massless
relativistic chiral (charge-density) fluctuations. This system has a conserved vector current, 
$J^{\mu}= \frac{e}{2\pi} \varepsilon^{\mu \nu} \partial_{\nu} \varphi$, and, in the ``relativistic'' 
approximation and for a vanishing external electric field, a conserved axial-vector current, 
$J^{\mu}_{5}$, as discussed in Section 2. In the presence of a non-vanishing external electric field, $E$, 
with vector potential $A_{\mu}$ the currents
$$\widehat{J}^{\,\mu}_{\ell/r} := J^{\mu}_{\ell/r} \mp \frac{Q}{2\pi} \varepsilon^{\mu \nu} A_{\nu}$$
are conserved, ($\partial_{\mu} \widehat{J}^{\mu}_{\ell/r} =0$), but \textit{not} gauge-invariant. 
However, the chiral charges
\begin{equation}\label{chiral charges}
N_{\ell/r}:= \int \widehat{J}^{\,0}_{\ell/r}(t, \underline{x}) d \underline{x}
\end{equation}
are \textit{not only conserved}, but also \textit{gauge-invariant}. It is important to observe (see \cite{ACF}) 
that if expressed in terms of the scalar field $\varphi$ introduced in \eqref{current potential} and 
\eqref{axialcurrent}, these charges are \textit{independent} of the parameters describing the 
interactions in the Tomonaga-Luttinger liquid \cite{Tomonaga, Luttinger}.

Let $H$ denote the Hamiltonian of the electron gas. 
The equilibrium state of this system at inverse temperature $\beta$ is given by the density matrix
\begin{equation}\label{densitymatrix}
P_{\mu_{\ell}, \mu_{r}}:= \Xi_{\beta, \mu_{\ell}, \mu_{r}}^{-1} \text{exp}(-\beta H_{\mu_{\ell}, \mu_{r}})\,,
\end{equation}
where $\Xi_{\beta, \mu_{\ell}, \mu_{r}}$ is the grand-canonical partition function, $\mu_{\ell}$ and $\mu_{r}$ denote the chemical potentials of the electron reservoirs on the right end of the wire (injecting left-moving electrons into the wire) 
and on the left end of the wire, respectively, and
 $$H_{\mu_{\ell}, \mu_{r}}:= H-\mu_{\ell} N_{\ell} - \mu_{r}N_{r}.$$
Expectations with respect to $P_{\mu_{\ell}, \mu_{r}}$ are denoted by $\langle (\cdot) \rangle_{\mu_{\ell}, \mu_{r}}$.
The electric current, $I$, flowing through the wire in the state $\langle (\cdot)\rangle_{\mu_{\ell}, \mu_r}$ 
is given by
\begin{align}\label{I}
I:= \langle J^{1}(x) \rangle_{\mu_{\ell}, \mu_{r}} = -\frac{e}{2 \pi} \langle \frac{\partial \varphi(x)}{\partial t} \rangle_{\mu_{\ell}, \mu_{r}} = -i\frac{e}{2\pi}\langle [H, \varphi(x)] \rangle_{\mu_{\ell}, \mu_{r}} \,,
\end{align}
where the last equation expresses the \textit{Heisenberg equation of motion}. Formally, one has that
\begin{equation}\label{commutator}
\langle [H, \varphi(x)] \rangle_{\mu_{\ell}, \mu_{r}} = \langle H\,\varphi(x) - \varphi(x)\,H\rangle_{\mu_{\ell}, \mu_r}\,.
\end{equation}
Since the charges $N_{\ell}$ and $N_r$ are conserved, one is tempted to conclude that the right side of
this equation vanishes and, hence, that $I=0$. This conclusion turns out to be \textit{wrong,} because 
the expression on the right of \eqref{commutator} is meaningless, due to non-trivial zero-modes of 
the field $\varphi$ that can be present because $\varphi$ is massless. In order to arrive at expressions 
that are meaningful, we replace the massless scalar field $\varphi$ used in the bosonization of the electron 
gas in the wire by a scalar field with a small but \textit{positive} mass that, at the end of our calculations, 
we will let approach 0. As long as this regularizing mass does not vanish, the Hamiltonian $H$ does \textit{not} 
commute with the charges $N_{\ell}$ and $N_r$, and hence \eqref{commutator} need \textit{not} vanish. 
To find the correct expression for the current $I$ passing through the wire, we use that
 \begin{align}\label{conductance}
I &\overset{\eqref{commutator}}{=}  -i\frac{e}{2\pi}\langle [H, \varphi(x)] \rangle_{\mu_{\ell}, \mu_{r}} \nonumber \\
 &\,\,= -i\frac{e}{2\pi} \langle [H_{\mu_{\ell}, \mu_{r}}, \varphi(x)] + [\mu_{\ell} N_{\ell} + \mu_{r} N_{r}, \varphi(x)] \rangle_{\mu_{\ell}, \mu_{r}}
\end{align}
As long as the mass of $\varphi$ is positive we have that
$$\langle [H_{\mu_{\ell}, \mu_r}, \varphi(x)] \rangle_{\mu_{\ell}, \mu_{r}} = \langle H_{\mu_{\ell}, \mu_r}\,\varphi(x) - 
\varphi(x)\, H_{\mu_{\ell}, \mu_r}\rangle_{\mu_{\ell}, \mu_r},$$
and the right side of this identity is meaningful and \textit{does} vanish, as follows from \eqref{densitymatrix} 
and the cyclicity of the trace.
Using Eq.~\eqref{chiral charges} and the anomalous commutator \eqref{ACC}, we find that the remaining 
terms on the right side of expression \eqref{conductance} for the current $I$ add up to
\begin{align}\label{current} 
I&= - \frac{ie^{2}}{2 \pi} (\mu_{\ell}-\mu_{r}) \int \langle [\varpi(t, \underline{y}), \varphi(t, \underline{x})] \rangle_{\mu_{\ell}, \mu_{r}} d\underline{y} \nonumber\\
&= -\frac{e^{2}}{2\pi} (\mu_{\ell} - \mu_{r})\,, 
\end{align}
as follows from the canonical commutation relations between $\varpi$ and $\varphi$. Equation \eqref{current} remains
true, as the regularizing mass of $\varphi$ tends to 0.\\
Notice that $-(\mu_{\ell} - \mu_{r})=: \Delta V$ is the voltage drop through the wire. Re-installing Planck's constant $\hbar$, 
we find that 
$$I= \sigma\, \Delta V, \qquad \text{with  }\,\,  \sigma = \frac{e^{2}}{2\pi \hbar}\,.$$
Since electrons have spin $\frac{1}{2}$, there are actually \textit{two} species of charged particles 
(electrons with ``spin-up'' and ``spin-down'') per non-empty band in the wire. Thus, 
 $$I = 2n \frac{e^{2}}{h} \Delta V\,, \quad \text{for a wire with $n$ non-empty bands.}$$
This equation implies that the conductance $\sigma$ in a quantum wire is quantized in even multiples
of $e^{2}/h$.

For more general results on transport in 1D systems exhibiting conformal symmetry see \cite{Gawedzki} 
and references given there.
 
\section{The 2D quantum Hall effect}
In this section we outline some important elements of the \textit{theory of the quantum Hall effect} (QHE)
(see \cite{Prange} and references given there).
We consider the idealized set-up, sketched in Figure 1, below, for measurements of the Hall conductivity 
(= Hall conductance) and the longitudinal resistance (or conductance) of a two-dimensional electron gas 
formed at the interface of a semi-conductor and an insulator when a gate voltage perpendicular to the 
interface is turned on.
The two-dimensional electron gas is assumed to be confined to a compact region, $\Omega$, contained in the $xy$-plane 
and is subjected to a magnetic field $\vec{B}_{0}\,\perp\, \Omega$ (parallel to the $z$-axis). To observe the quantization 
of the Hall conductivity, $\sigma_H$, the filling factor $\nu:= n_{e}/(eB_{0}/h)$ of the electron gas (where $n_e$ 
is the density of the gas, and $B_0 := \vert \vec{B}_{0} \vert$) must be chosen in such a way that the longitudinal 
resistance, $R_L$, of the electron gas vanishes. (If the Hall (transverse) conductivity, $\sigma_H$, does not 
vanish then $R_L=0$ is equivalent to a vanishing longitudinal conductance.) To show, mathematically, that, 
for an \textit{interacting} two-dimensional electron gas in a uniform external magnetic field, one can choose 
the filling factor $\nu$ in such a way that $R_L = 0$ appears to be a very hard problem of quantum many-body 
theory that has not been solved under realistic assumptions about the electron gas but has been extensively 
studied numerically (see, e.g., \cite{Morf}). 
In the following, we attempt to solve a much easier problem: 
\textit{Assuming} that $R_L =0$, what can one say about the properties of the elctron gas and, 
in particular, about the possible values of the Hall conductivity $\sigma_H$ (under the condition that the diameter 
of $\Omega$ is much larger than the magnetic length)? This question has a very elegant, general answer (see
\cite{FS-RMP, Les Houches 94, ICM 94, FPSW}). In order to find this answer, we have to study the response 
of a 2D electron gas confined to $\Omega$ to a small perturbing electromagnetic field 
$(\vec{E}, \vec{B})$, with $\vec{E}\,\Vert\, \Omega$ and $\vec{B} \perp \Omega$. 
We set
$$\vec{B}^{tot}:= \vec{B}_0+\vec{B}, \quad B:= \vert \vec{B} \vert, \quad \underline{E}:=(E_1,E_2).$$

It is helpful to review the electrodynamics of 2D ``incompressible'' ($R_L =0$) electron gases.\,
The electromagnetic field tensor, $F=\big(F_{\mu\nu}\big)= \big(\partial_{\mu}A_{\nu} - \partial_{\nu}A_{\mu}\big), \,
\mu, \nu= 0,1,2,$\, in $2+1$ space-time dimensions is given by
 \begin{displaymath}
 {F}=  
 \left( 
 \begin{array}{ccc}
 0&E_{1}&E_{2}\\
 -E_{1}&0&-B\\
 -E_{2}&B&0 
 \end{array}
 \right)
\end{displaymath}
The electric current density, $j^{\mu}$, in the gas is given by
$$j^{\mu}(x):= \langle J^{\mu}(x) \rangle_{A}, \quad \mu=0,1,2\,,$$
where $\langle(\cdot)\rangle_A$ is the (time-dependent) state of the gas in a very slowly varying external
electromagnetic field with vector potential $A=(A_{\mu})$.
\begin{center}
\includegraphics[width=15.5cm]{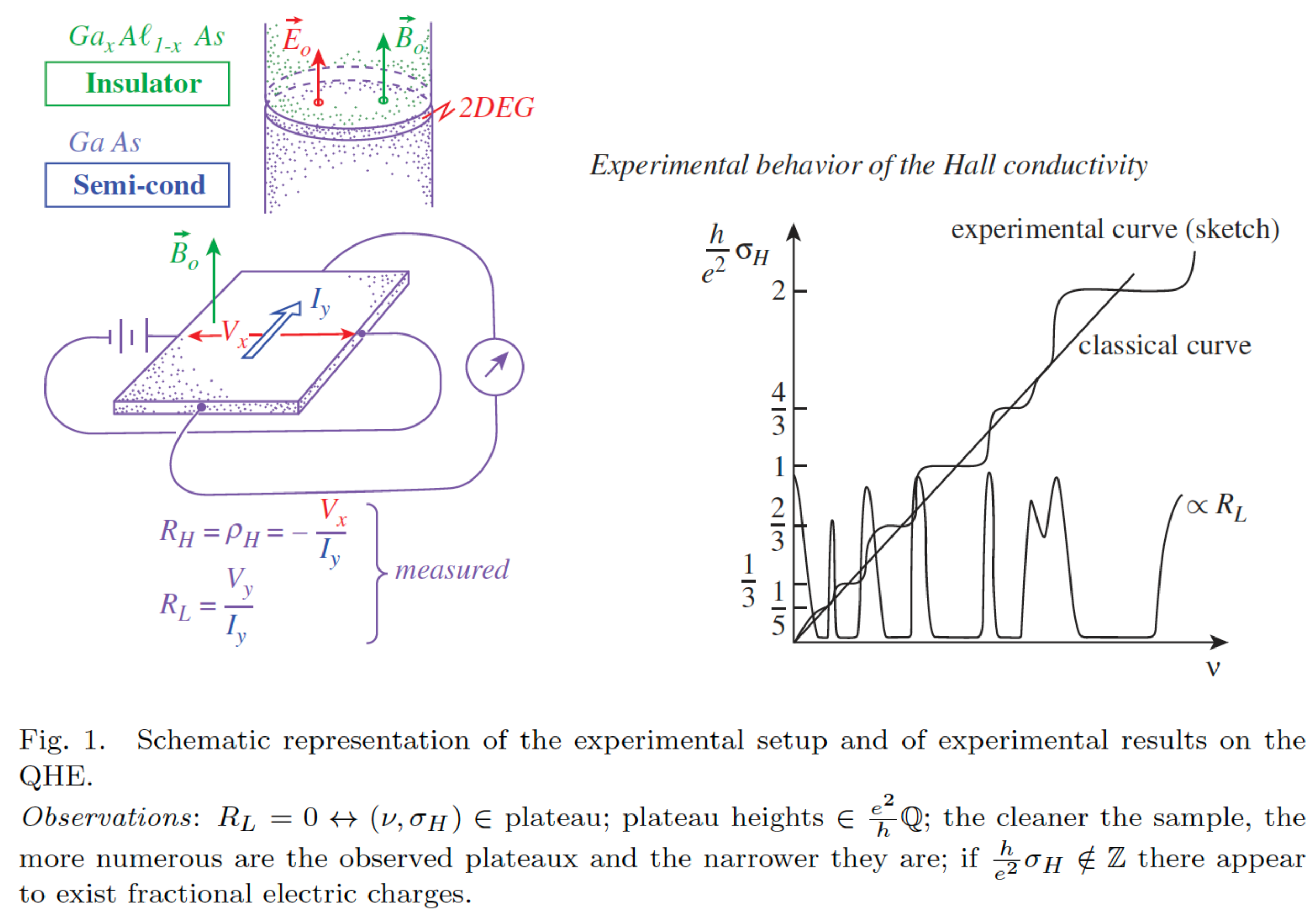} 
\end{center}
\hspace{0.7cm}{\small{\textit{Note:} Figures 1 through 3 have previously appeared in \cite{Faddeev}, figure 3 also in \cite{Les Houches 94}.}}
\subsection{The electrodynamics of 2D incompressible electron liquids -- anomalous chiral edge currents}
The basic equations of the $(2+1)$-dimensional electrodynamics of 2D incompressible electron gases (in units
where the speed of light $c=1$) are as follows.
\begin{enumerate}
\item[(i)] {\textbf{Hall's Law} -- (a phenomenological law)
 \begin{equation}\label{Hall law}
 j^{k}(x)= \sigma_{H} \varepsilon^{k\ell}E_{\ell}(x)\,, \,\,\, k, \ell = 1,2, \quad \text{assuming that }\,\,R_{L}=0,
 \end{equation}
 where $\varepsilon^{12}=-\varepsilon^{21}= 1, \varepsilon^{ii}=0$, which shows that if $\sigma_H \not= 0$ 
 parity ($P$) and time reversal ($T$) are broken. (Of course these symmetries are broken by the presence 
 of an external magnetic field.)}
 \item[(ii)]{\textbf{Charge conservation} -- (a fundamental law)
 \begin{equation}\label{continuity eq}
 \frac{\partial}{\partial t} \rho(x) + \underline{\nabla} \cdot \underline{j}(x) =0 \,.
 \end{equation}
 }
 \item[(iii)]{\textbf{Faraday's induction law} -- (a fundamental law)
 \begin{equation}\label{Faraday}
 \frac{\partial}{\partial t} B_{3}^{tot}(x) + \underline{\nabla} \wedge \underline{E}(x)=0 \,.
  \end{equation}
  Combining Eqs.~(i) through (iii), we obtain
  \begin{equation}\label{Rhodot}
  \frac{\partial}{\partial t}\rho \overset{(ii)}{=} -\underline{\nabla}\cdot \underline{j} \overset{(i)}{=} -\sigma_{H} \underline{\nabla} \wedge \underline{E} \overset{(iii)}{=} \sigma_{H} \frac{\partial}{\partial t} B\,.
 \end{equation}}
\item[(iv)]{\textbf{Chern-Simons Gauss law} -- (useful to determine the Hall conductivity)\\
Integrating \eqref{Rhodot} in time $t$, with integration constants chosen as follows
 $$j^{0}(x):= \rho(x) + e\cdot n_e, \quad B(x)= B_{3}^{tot}(x)- B_{0}\,,$$
 we find the so-called \textit{Chern-Simons Gauss law} (see eqs.~\eqref{general Hall} and \eqref{CS action}, below)
 \begin{equation}\label{CS-Gauss}
 j^{0}(x)=\sigma_{H} B(x)\,.
 \end{equation}
 This equation is also known under the name of \textit{Streda formula}; see \cite{Streda}.
Integrating both sides of Eq. \eqref{Rhodot} in time from $t=t_i$ to $t=t_f$ and over sample space $\Omega$, 
we conclude that
\begin{equation}\label{charge transport}
\Delta Q= \sigma_H \Delta \Phi,
\end{equation}
where $\Delta Q$ is the change of the electric charge stored in the sample and $\Delta \Phi$ 
is the change of the magnetic flux through the sample during the time interval $[t_i,t_f]$.
 }
 \end{enumerate}
\textbf{Remark.}
For 2D incompressible electron liquids of \textit{non-interacting} electrons moving in a disordered potential landscape, 
the Chern-Simons Gauss law \eqref{charge transport} has been invoked in calculations of the Hall conductivity, 
$\sigma_H$, with the purpose to exhibit its topological character (see \cite{Laughlin, ASS}). 
But this law, as well as Hall's law \eqref{Hall law} can equally well be used to calculate the Hall 
conductivity for incompressible \textit{interacting} 2D electron liquids and to exhibit the topological nature 
of $\sigma_H$. For example, one may consider an electon liquid confined to a 2D torus, $\mathcal{T}$, 
and thread a magnetic flux tube through the interior of $\mathcal{T}$ that does not cross/intersect $\mathcal{T}$ and
is parallel to a non-contractible cycle, $\gamma$, of $\mathcal{T}$.\footnote{Instead, we could consider a very
large Corbino disk and study charge transport from the inner to the outer boundary circle when the magnetic 
flux threaded through its hole changes.} 
One may then ask how much electric charge, $\Delta Q_{\widetilde{\gamma}}$, crosses a cycle, 
$\widetilde{\gamma}$, on $\mathcal{T}$ conjugate to $\gamma$ (i.e., $\gamma$ and $\widetilde{\gamma}$ intersecting each other in a single point of $\mathcal{T}$) when 
the magnetic flux in the interior of the torus (parallel to $\gamma$) changes by an amount $\Delta \Phi$. 
A combination of Eq.~\eqref{Hall law} with Faraday's induction law \eqref{Faraday} shows that the quotient 
$\Delta Q_{\widetilde{\gamma}}/\Delta \Phi$ is given by $\sigma_H$. Assuming that the electric charge of all charge carriers in 
the system is an integer multiple of the elementary electric charge $e$, and assuming that the 
longitudinal conductance vanishes, one can invoke the theory of the Aharonov-Bohm effect to 
conclude that the state of this system is unchanged if the magnetic flux in the interior of 
$\mathcal{T}$ is slowly increased by an integer multiple of the flux quantum $h/e$. This implies
that, in the process of increasing $\Delta \Phi$ by $h/e$, an integer number of charge carriers with a total electric charge 
$Ne, N\in \mathbb{Z},$ must cross the cycle $\widetilde{\gamma}$. Hence
$$ \Delta Q_{\widetilde{\gamma}} = N e =\sigma_H (h/e) \quad \Leftrightarrow \quad \sigma_H = 
N \frac{e^{2}}{h}\,, \,\,N\in \mathbb{Z}\,.$$
These considerations do not explain why the ``Hall fraction,'' $\frac{h}{e^{2}}\sigma_H$, is often observed \textit{not} to be
an integer, but rather a \textit{rational number} (most often with an odd denominator). It is plausible to imagine that 
if the Hall fraction is not an integer then there must exist \textit{fractionally charged quasi-particles}, as originally 
suggested by Tsui and Laughlin (see Subsection 4.1).

The sign of $\sigma_H$ is determined by the nature of the quasi-particles carrying the Hall current, 
namely whether these quasi-particles are electrons or holes. (Historically, this fact led to the discovery 
of holes in a nearly full conduction band of semi-conductors.)
 
Hall's law \eqref{Hall law} and the Chern-Simons Gauss law \eqref{CS-Gauss} yield the equation
\begin{equation}\label{general Hall}
 j^{\mu}(x)= \sigma_{H} \varepsilon^{\mu\nu\lambda}F_{\nu\lambda}(x)\,,
\end{equation}
which is a \textit{generally covariant} relation between the current density $j^{\mu}$ of the electron gas and the electormagnetic
field tensor $F_{\nu \lambda}$.

Equation \eqref{general Hall} evokes to the following puzzle.
 \begin{equation}\label{puzzle}
 0\overset{(ii)}{=} \partial_{\mu}j^{\mu} \overset{(iii), \eqref{general Hall}}{=} \varepsilon^{\mu\nu\lambda}(\partial_{\mu} \sigma_{H}) F_{\nu\lambda}\,.
\end{equation}
We observe that, while the left side of this equation vanishes by charge conservation, the right side does 
\textit{not} vanish wherever the value of $\sigma_{H}$ jumps, 
as, for example, at the boundary, $\partial \Omega$, of the sample region $\Omega$. So, what is going on?\\

\noindent
\textbf{Solution of Puzzle}: 
In equation \eqref{general Hall} $j^{\mu}$ is the \textit{bulk} current density, $j^{\mu}_{bulk}$, which is apparently \textbf{not} 
conserved! The \textit{conserved total electric current diensity,} $j^{\mu}_{tot}$, can be decomposed into
 \begin{equation}\label{edge currents}
 j^{\mu}_{tot}= j^{\mu}_{bulk} + j^{\mu}_{edge}, 
 \end{equation}
 with
 $$\partial_{\mu}j^{\mu}_{tot}=0, \,\,\,\text{   but   }\,\,\, \partial_{\mu}j^{\mu}_{bulk} \overset{\eqref{puzzle}}{\not=} 0\,.$$
We note that the support of the edge current density $j^{\mu}_{edge}$ coincides with the support of 
$\underline{\nabla} \sigma_H$,
 $$\text{supp }j^{\mu}_{edge} = \text{supp}\lbrace \underline{\nabla} \sigma_{H} \rbrace \supseteq \partial \Omega, \qquad 
\text{with }\quad\,\, \underline{j}_{edge} \perp \underline{\nabla} \sigma_{H}\,.$$

Combining \eqref{puzzle} (setting $j^{\mu} = j^{\mu}_{bulk}$!) with \eqref{edge currents}, we find that
\begin{equation*}
\partial_{\mu} j^{\mu}_{edge} \overset{\eqref{edge currents}}{=} -\partial_{\mu} j^{\mu}_{bulk}\vert_{\text{supp}\lbrace \underline{\nabla} \sigma_{H}\rbrace} \overset {\eqref{general Hall}}{=} - \varepsilon^{\mu\nu} \partial_{\mu}\sigma_{H} 
E_{\nu}\,.
\end{equation*}
When restricted to the support of $\underline{\nabla}\sigma_H$ this equation becomes
\begin{equation}\label{2D anomaly}
\partial_{\mu} j^{\mu}_{edge} = -(\Delta \sigma_H) E_{\Vert}\Vert_{\text{supp}\{\underline{\nabla}\sigma_H\}}\,,
\end{equation}
where $\Delta \sigma_H$ is the jump of the Hall conductivity across the edge in question. In the following, 
we assume that $\sigma_H$ is constant throughout the interior of $\Omega$ and vanishes outside $\Omega$ 
(so that $\Delta \sigma_H = \sigma_H$ in \eqref{2D anomaly}). 
This is not a realistic assumption; but it simplifies our discussion without introducing any misleading reasoning.
Equation \eqref{2D anomaly} shows that $j^{\mu}_{edge}$ is an anomalous (chiral) current density in $1+1$ dimensions. 
This equation can be viewed as a simple manifestation of ``holography'': The edge displays as much 
information about the Hall conductivity of the system as the bulk. \\
Here is a classical-physics argument determining the chirality of $j^{\mu}_{edge}$: At the edge of the sample the 
Lorentz force acting on electrons must be cancelled by the force confining them to the interior of the sample. Thus,
$$\frac{e}{c} B^{tot} v_{\Vert}^{k} = \varepsilon^{k\ell}\frac{\partial V_{edge}}{\partial x^{\ell}}\,,$$
where $V_{edge}$ is the potential of the force confining electrons to the interior of the sample region 
$\Omega$. This enables us to find the equation for the chiral motion of degrees of freedom localized at
the boundary, $\partial \Omega$, of the system; (``skipping orbits'').
\begin{center}
\includegraphics[width= 15cm]{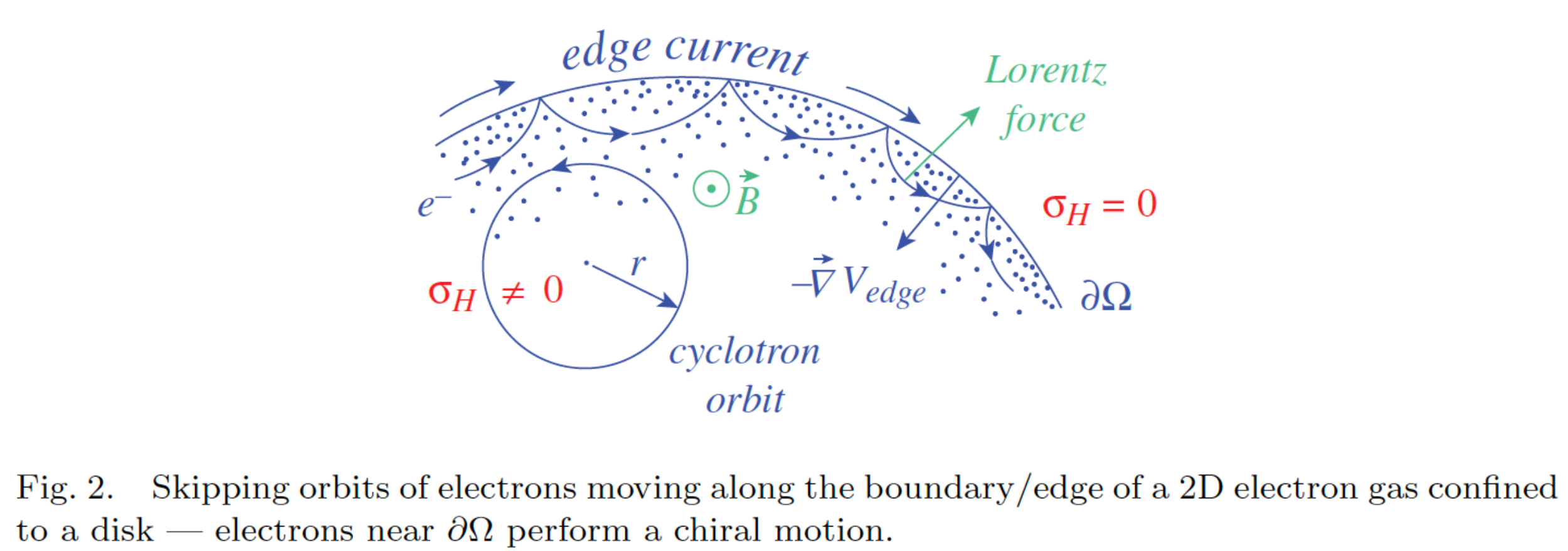}
\end{center}
From the theory of the chiral anomaly in $1+1$ dimensions one may infer that, at the boundary $\partial \Omega$,
\begin{equation}\label{Hall cond}
\partial_{\mu} j^{\mu}_{edge} = - \frac{e^{2}}{h} \big( \sum_{\text{species } \alpha} Q_{\alpha}^{2}\big) E_{\Vert} \overset{\text{with  }\eqref{2D anomaly}}{\Rightarrow}
 \sigma_{H}= \frac{e^{2}}{h} \sum_{\text{species  }\alpha} Q_{\alpha}^{2}
\end{equation}
where $eQ_{\alpha}$ is the ``charge'' (see \eqref{current potential}) of the edge current, $j^{\mu}_{\alpha}$, 
corresponding to species $\alpha$ of clockwise-chiral edge modes; (similar contributions will come from 
counter-clockwise chiral modes, but with \textit{reversed} sign).\\
Apparently, we have re-discovered \textit{Halperin}'s chiral edge currents \cite{Halperin}, but only using very general 
arguments that are valid even for interacting electron gases.
We observe that if $\sigma_{H} \not\in \frac{e^{2}}{h} \mathbb{Z}$ then there exist chiral currents associated with
\textit{fractionally charged} modes propagating along the edges.

A phenomenon in classical physics analogous to the anomalous edge currents of 2D electron gases is found in the 
physics of hurricanes:
$$\vec{B} \rightarrow \vec{\omega}_{earth}, \text{  Lorentz force } \rightarrow \text{  Coriolis force  }, V_{edge} 
\rightarrow \text{  air pressure  }.$$
This shows (among other things) that hurricanes on the northern hemisphere always turn anti-clockwise. 
Related phenomena are the equatorial winds and chiral coastal water waves, called \textit{Kelvin waves} 
(after Lord Kelvin, who discovered them). The take-home message of this observation is that anomalies, 
anomaly inflow and anomaly cancellation are effects that already appear in classical physics, in particular 
in the aerodynamics of the Earth's atmosphere and in the dynamics of fluids in non-inertially moving containers, 
such as the Earth's oceans.

Next, we determine the (bulk- and edge) effective action of 2D incompressible electron gases exhibiting the 
quantum Hall effect.
As above, we consider a 2D electron gas in a neutralizing ionic background subject to a constant transversal 
magnetic field $\vec{B}_{0}$. The electrons are confined to a region $\Omega$ in the $xy$-plane. The space-time 
of the system is given by $\Lambda=\Omega \times \mathbb{R}$ (with $\mathbb{R}$ identified with the time axis). 
We suppose that the electrons are coupled to an external electromagnetic field with 
vector potential $A = A_0\, \text{d}t + A_1\, \text{d}x + A_2\, \text{d}y$ 
describing a small, slowly varying perturbing electromagnetic field $(\underline{E}, B)$.\\
We assume that the 2D electron gas is an insulator, i.e., that the longitudinal conductance vanishes. 
It is then easy to determine the form of the effective action, $S_{eff}(A)$, of this system as a functional 
of the external vector potential $A$ in the limiting regime of very large distances and very low frequencies,
i.e., in the scaling limit; as alluded to in Section 1 (see \cite{FS-RMP, Les Houches 94} for details):
 \begin{align}\label{eff act}
 S_{eff}(A) = & \frac{\sigma_{H}}{2}\int_{\Lambda} A\wedge [dA+K] \,+ \text{ boundary action}\nonumber \\ 
  + & \frac{1}{2}\int_{\Lambda} \text{d}^{3}x \{ \underline{E}(x)\cdot \varepsilon \underline{E}(x) - \mu^{-1} B^{2}(x) \} + ... ,
  \end{align}
where the first term on the right side is the topological Chern-Simons action (which is ``marginal'' in the infrared), 
and its coefficient, $\sigma_H$, turns out to be the \textit{Hall conductivity}, $K$ is the Gauss curvature 2-form 
on the sample space $\Omega$, which gives rise to the famous ``shift''; $\varepsilon=$ tensor of dielectric 
constants, $\mu=$ magnetic permeability (and the last term on the right side of \eqref{eff act} is 
``irrelevant'' in the infrared).
 
 The presence of the Chern-Simons action on the right side of \eqref{eff act} can also be inferred from 
 Eq.~\eqref{general Hall}.
 Omitting curvature terms ($K=0$),
  \begin{align}\label{bulk action}
 j^{\mu}_{bulk}=& \langle J^{\mu}(x) \rangle_{A} \equiv \frac{\delta S_{\Lambda}(A)}{\delta A_{\mu}(x)} \nonumber\\
 \overset{\eqref{general Hall}}{=}& \sigma_{H} \,\varepsilon^{\mu\nu\lambda}F_{\nu\lambda}(x), \quad \text{for  }\,\,x\not\in \partial\Lambda\,.
 \end{align}
 Integrating Eq.~\eqref{bulk action}, we find that the effective action $S_{eff}(A)\equiv S_{\Lambda}(A)$ is given by
 \begin{equation}\label{CS action}
  S_{\Lambda}(A)= \frac{\sigma_{H}}{2}\int_{\Lambda} A\wedge dA + \text{ boundary action}\,.
 \end{equation}
That there \textit{must} be a boundary action, denoted in the following by $\Gamma_{edge}(a)$, is a consequence 
of the fact that the Chern-Simons bulk action is \textit{not} gauge-invariant on a space-time $\Lambda$ with a 
non-empty boundary $\partial \Lambda$: Under a gauge transformation 
$A_{\mu} \rightarrow A_{\mu} + \partial_{\mu} \chi$, the Chern-Simons action changes by a boundary term
 \begin{equation}\label{boundary anomaly}
\frac{\sigma_{H}}{2} \int_{\partial \Lambda} [\chi dA]\vert_{\partial \Lambda}
\end{equation}
 This anomaly must be cancelled by the anomaly of a boundary action.

Returning to Eqs.~\eqref{2D anomaly} and\eqref{Hall cond}, we guess that the boundary action must be the 
generating functional of the connected Green functions of the (quantum-mechanical) anomalous chiral edge 
current density
$$\frak{j}^{\mu}_{edge}= \sum_{\alpha} \frak{j}^{\mu}_{\alpha} \,, \quad \alpha=1,2, ... ,$$ 
where $\alpha$ labels the different species of charged chiral edge modes. 
The charge of the chiral edge modes corresponding to $\frak{j}^{\mu}_{\alpha}$ has been denoted by $eQ_{\alpha}$.
Let $v_{\alpha}$ denote the propagation speed of the chiral modes that give rise to the edge current density 
$\frak{j}^{\mu}_{\alpha}$. This propagation speed (which in general depends on the species $\alpha$) plays 
the role of the ``speed of light'' in 2D current algebra. We introduce ``light-cone coordinates'', 
$u^{+}, u^{-}$, on $\partial \Lambda$. Let $a \equiv a_{+}du^{+} + a_{-}du^{-}:=A_{\Vert}$ denote the electromagnetic 
vector potential (which is a 1-form) restricted to the $1+1$-dimensional boundary, $\partial \Lambda$, of the 
\mbox{$(2+1)$-dimensional} space-time. \\
The effective action of the chiral edge current density $\frak{j}^{\mu}_{\alpha}$ is given by
\begin{equation}\label{boundary action}
\frac{(eQ_{\alpha})^{2}}{h}  \Gamma_{\partial \Lambda}^{(\alpha)}(a)\,, \text{ with}\quad  \Gamma_{\partial \Lambda}^{(\cdot)}(a):= \frac{1}{2} \int_{\partial \Lambda} [a_{+}a_{-} - a_{\pm} \frac{\partial^{2}_{\mp}}{\Box} a_{\pm}] du^{+}\,du^{-}\,,
 \end{equation}
 where $\Box:=\partial_{+}\partial_{-}$, and the choice of subscripts ``$+$'' and ``$-$'' in the last term on the right side of 
 \eqref{boundary action} depends on the chirality of the modes giving rise to the edge current $\frak{j}^{\mu}_{\alpha}$; 
 (the dependence of $\Gamma_{\partial \Lambda}^{(\alpha)}(a)$ on $\alpha$ is determined by the chirality 
 of the mode $\alpha$ and its propagation speed $v_{\alpha}$).

The reader is invited to verify that the anomaly \eqref{boundary anomaly} of the bulk effective action, namely the term 
$\frac{\sigma_{H}}{2} \int_{\partial \Lambda} [\chi dA]\vert_{\partial \Lambda}$, is cancelled by 
the anomaly of the edge effective action, 
\begin{equation}\label{edge act}
\Gamma_{edge} (a):=-\sum_{\text{species } \alpha} \frac{(eQ_{\alpha})^{2}}{h}\Gamma_{\partial \Lambda}^{(\alpha)}(a)\,,
\end{equation}
under a gauge transformation $a \rightarrow a+d\chi\vert_{\partial \Lambda}$ if and only if 
\begin{equation}\label{edge=bulk}
\partial_{\mu}\, \big(\delta\,\Gamma_{edge}(a)/\delta a_{\mu}(x)\big) = \partial_{\mu}j^{\mu}_{edge}(x)\equiv
- \sigma_{H}^{edge}E_{\Vert} \overset{!}{=} - \sigma_{H} E_{\Vert}\,,
\end{equation}
see \eqref{2D anomaly}; whence
\begin{equation}\label{edge=bulk}
\sigma_{H}^{edge} \overset{\eqref{edge act}}{=} \frac{e^{2}}{h} \sum_{\alpha} Q_{\alpha}^{2} = \sigma_H\,.
\end{equation}
This shows that edge- and bulk Hall conductivities must \textit{coincide}, as already emphasized in \cite{FS-RMP, 
Les Houches 94}.
Merely for simplicity, we have assumed here (and will continue to assume in the following) that all edge modes have the 
\textit{same} chirality; otherwise, appropriate signs have to be introduced in these formulae.\\

Whatever has been said here about Hall insulators also applies to so-called \textit{Chern insulators}, 
which break reflection- and time-reversal invariance \textit{even} in the absence of an external magnetic field; 
e.g., because the bulk of the material is doped by magnetic impurities; (as in the Haldane model \cite{Haldane};
see also \cite{F-Kerler}).

\subsection{Classification of ``abelian'' Hall fluids} 
In this subsection I sketch a general classification of 2D insulators with broken parity and time reversal 
invariance (Hall- or Chern insulators) exhibiting quasi-particles with \textit{abelian braid statistics}.\footnote{States 
exhibiting quasi-particles with \textit{non-abelian braid statistics} are discussed in my work \cite{FPSW} with 
B.~Pedrini, Chr.~Schweigert and J.~Walcher.} I will describe these systems in the limit of very large distance 
scales (i.e., I consider very large samples in units of the magnetic length) and of very low frequencies; 
i.e., in the infrared \textit{scaling limit}. Units will be used such that $\frac{e^{2}}{\hbar}=1$. Let $J$ denote 
the total operator-valued electric current density (bulk- plus edge current density), which is conserved: 
$\partial_{\mu}J^{\mu} =0$. \\

\noindent
\textbf{Ansatz:}
\begin{equation}\label{4.14}
J= \sum_{\alpha=1}^{N} Q_{\alpha} J_{\alpha},
\end{equation}
where the currents $J_{\alpha}$ are assumed to be canonically normalized and conserved, with charges 
$Q_{\alpha}\in\mathbb{R}$.\vspace{0.2cm}\\
\noindent
\textbf{Comment.} This ansatz has been criticized, a common objection being that there is no reason why 
there should be more than \textit{one} conserved current density, namely the total electric current
density. To counter this objection, I first consider a gas of electrons confined to a planar region $\Omega$ 
with the topology of a disk and a smooth boundary; a uniform external magnetic field transversal to
$\Omega$ is turned on, and it is assumed that there are no electron-electron interactions. This
system can be understood by studying the one-electron Hamiltonian: its eigenvalues form Landau
levels. The eigenvalues corresponding to eigenfunctions localized close to the boundary, $\partial \Omega$, 
of $\Omega$, are shifted upwards to rather high energies by the boundary potential that confines the electrons 
to the interior of $\Omega$ (a fact that can be verified in perturbation theory); i.e., the Landau levels
end in bands of energies corresponding to chiral modes with eigenfunctions localized near $\partial \Omega$ 
and extended along $\partial \Omega$. The \textit{ground-state} of the electron gas is obtained by filling \textit{all} 
the eigenstates corresponding to energy levels below the Fermi energy, $E_F$. Assuming that $E_F$ is such 
that a certain number, $n$, of Landau levels are partially filled then there are $n$ energy bands corresponding 
to chiral boundary modes that are partially filled up to an energy $\leq E_F$. The occupied bulk and boundary 
states then carry $n$ currents that, at very low frequencies, are \textit{separately conserved.} 

If the electrons move in a disordered potential landscape the energy eigenvalues of the Landau levels
are spread out in ``Landau bands.'' Eigenvalues close to the Landau-band edges correspond to localized
eigenfunctions. But in the middle of each Landau band there are eigenvalues corresponding to \textit{extended}
states. The eigenvalues corresponding to eigenfunctions located near $\partial \Omega$ end in an energy band 
of chiral boundary modes that is close to the band found when the disorder potential vanishes (as follows from 
arguments similar to Halperin's \cite{Halperin}). By dilating the region $\Omega$ one may approach the 
\textit{scaling limit}. In this limit, the extended states in the middle of each filled Landau band, together with
the corresponding states of chiral boundary modes with energies $\leq E_F$, then carry a current that, at 
the very low frequencies that survive in the scaling limit, is \textit{conserved.}

Thus, for non-interacting electron liquids, our ansatz appears to be justified in the scaling limit. (For finitely
extended samples, there are corrections possibly describing some mixing of the currents, which, however,
vanish in the scaling limit. I realize that mathematical rigor would require a more thorough analysis.)

For interacting electron liquids, the success of the \textbf{Ansatz}, which will be demonstrated below, 
appears to justify it.\\

\textit{\underline{A topological field theory of conserved currents emerging in the scaling limit}:}
On a 3D space-time $\Lambda= \Omega \times \mathbb{R}$, a conserved current 
$J$ can be derived from a vector potential, $B$. If $\iota$ denotes the 2-form dual to $J$ then the
continuity equation $\partial_{\mu} J^{\mu}=0$ is equivalent to $d\iota =0$. By Poincar\'e's lemma,
$$\iota=\frac{1}{\sqrt{2\pi}}dB,$$
where the vector potential $B$ is a 1-form, which is determined by $\iota$ up to the gradient of a scalar function, $\beta$:
 $B$ and $B+d\beta$ yield the \textit{same} 2-form $\iota$.\\

For a gapped 2D insulator, the effective field theory of the currents $\big(J_{\alpha}\big)_{\alpha=1}^{N}$ 
must be \textit{topological} in the scaling limit. If the symmetries of reflection in lines and time reversal 
are broken the ``most relevant'' term in the action functional of the potentials, 
$\underline{B}:=(B_{\alpha})_{\alpha=1}^{N}$, of the currents $J_{\alpha}$ is the \textit{Chern-Simons action}
\begin{equation}\label{4.15}
S_{\Lambda}(\underline{B},A):= \sum_{\alpha=1}^{N}\int_{\Lambda}\lbrace \frac{1}{2}B_{\alpha}\wedge dB_{\alpha} +A\wedge \frac{Q_{\alpha}}{\sqrt{2\pi}}dB_{\alpha}\rbrace +\text{boundary terms} +\dots,
\end{equation}
where $A$ is the electromagnetic vector potential, and the boundary terms must be added to cancel the anomalies 
of the bulk action under the ``gauge transformations''
$$B_{\alpha}\rightarrow B_{\alpha} +d\beta_{\alpha},\quad\,A\rightarrow A+d\chi\,.$$
Carrying out the oscillatory Gaussian integrals over the potentials $B_{\alpha}$, we find
\begin{equation}\label{4.16}
e^{iS_{\Lambda}(A)} \equiv \mathcal{Z}_{\Lambda}(A) := \int \text{exp}(iS_{\Lambda}\big(\underline{B},A)\big) \prod_{\alpha=1}^{N} \mathcal{D}B_{\alpha} = \text{exp}\left(i \frac{\sigma_H}{2}\int_{\Lambda}A\wedge dA+\Gamma_{edge}(A_{\Vert})\right),
\end{equation}
where 
$$\sigma_{H}= \frac{1}{2\pi}\sum_{\alpha=1}^{N} Q_{\alpha}^{2}\,,$$
by \eqref{4.15}; compare to \eqref{Hall cond}.

Physical states of the Chern-Simons theory with action as in \eqref{4.15} can be constructed from 
\textit{``Wilson networks''}, $W$, of Wilson lines and -flux tubes contained in the half space $\Lambda_{-}$, 
where $\Lambda_{+/-}:= \Omega\times \mathbb{R}_{+/-}= \Omega\times\{t\,\vert\, t \geq 0/t\leq 0\},$
whose open lines/flux tubes end in $\Omega_0 \equiv \Omega \times \{t=0\}$. 
Given such a network $W$, let $\vert W\rangle$ denote the physical state corresponding to $W$. 
Let $\Theta(W)$ denote the network contained in $\Lambda_{+}$ arising from $W$ by reflection in $\Omega_0$, 
followed by complex conjugation. If $W_{1}$ and $W_{2}$ are two such networks with the property that their 
intersections with $\Omega_0$, more precisely their \textit{fluxes} through $\Omega_0$, coincide we may 
consider the \textit{gauge-invariant} network, $W_{1}\circ \Theta(W_{2})$, arising by multiplying $W_{1}$ 
with $\Theta(W_{2})$; (graphically this amounts to concatenation at coinciding points/flux patches in $\Omega_0$). 
Then the scalar product of the state $\vert W_{1}\rangle$ with the state $\vert W_{2}\rangle$ is given by
\begin{equation}\label{4.17}
\langle W_{2}\Vert W_{1}\rangle := \mathcal{Z}_{\Lambda}(A)^{-1} \int \big(W_{1}\circ \Theta(W_{2})\big)(\underline{B})\, \text{exp}(iS_{\Lambda}\big(\underline{B},A)\big) \prod_{\alpha=1}^{N} \mathcal{D}B_{\alpha}\,.
\end{equation}
\textbf{Fact:} The Hamiltonian of a Hall insulator described by \eqref{4.15} through 
\eqref{4.17} vanishes. Thus, in the scaling limit, all excitations are \textit{``static''}.

The operator, $\mathcal{Q}_{\mathcal{O}}$, measuring the electric charge stored in states corresponding to Wilson 
networks whose supports intersect the sample space $\Omega_0$ in a region $\mathcal{O}$, can be constructed from
the Wilson loops
$$\text{exp}(i \varepsilon \mathcal{Q}_{\mathcal{O}}):=\text{exp}\big(i \varepsilon \int_{\mathcal{O}} J^{0}d^{2}x\big)=\text{exp}\big(i \sum_{\alpha=1}^{N} \varepsilon \frac{Q_{\alpha}}{\sqrt{2 \pi}} \int_{\partial{\mathcal{O}}}B_{\alpha} \big)\,, \quad \varepsilon\in \mathbb{R}\,.$$
Because the ground-state energy of a Hall insulator is separated from the rest of the energy spectrum by a strictly 
\textit{positive (mobility) gap}, \textit{electric charge} is a good quantum number to label its physical states at zero 
temperature. In other words, the charge operators 
$$ 
\mathcal{Q}_{\mathcal{O}}\,\,\,\text{ and  } \,\,\mathcal{Q}:= \text{lim}_{\mathcal{O}\nearrow \Omega} \mathcal{Q}_{\mathcal{O}}
$$
are well defined on (zero-temperature) physical states.\footnote{The same conclusion is reached by noticing that 
all Wilson loop expectations have perimeter decay and then invoking ``\,'tHooft duality''.}\\
The electric charges, $q_{\mathcal{O},1}$ and $q_{\mathcal{O},2}$, contained in a region $\mathcal{O}\subset \Omega_0$ of 
two states $\vert W_{1}\rangle$, $\vert W_{2}\rangle$ with the property that $W_{1}\circ \Theta(W_{2})$ is gauge-invariant 
must be \textit{identical}, i.e., $q_{\mathcal{O},1}=q_{\mathcal{O},2}\equiv q_{\mathcal{O}}$.
Their electric charge contained in $\mathcal{O}$, i.e., the number $q_{\mathcal{O}}$, is given by
\begin{align}\label{4.18}
\text{exp}(i\varepsilon q_{\mathcal{O}}) \langle W_{2}\Vert W_{1}\rangle = \mathcal{Z}_{\Lambda}(A)^{-1}
 \int \big(W_{1}\circ \Theta(W_{2})\big)(\underline{B})\,
\text{exp}(i\varepsilon \mathcal{Q_{\mathcal{O}}})\, \text{exp}(iS_{\Lambda}\big(\underline{B},A)\big) 
\prod_{\alpha=1}^{N} \mathcal{D}B_{\alpha}\,,
\end{align}
with $\varepsilon \in \mathbb{R}$ arbitrary. If a Wilson network $W$ creates a physical state $\vert W \rangle$ 
describing $n$ electrons or holes located inside a region $\mathcal{O}\subset \Omega_0$ when applied 
to the ground-state of a Hall insulator then the charge, $q_{\mathcal{O}}\equiv q_{\mathcal{O}}(W)$, 
contained in $\mathcal{O}$ is given by $q_{\mathcal{O}}(W) = -n+2k$, where $k$ is the number of holes in $\mathcal{O}$.
If $-n+2k$ is odd, i.e., if $n$ is \textit{odd}, then the excitation created by $W$ inside $\mathcal{O}$ must have 
Fermi-Dirac statistics, and if $n$ is \textit{even} it must have Bose-Einstein statistics; i.e., there is a \textit{connection
between electric charge and quantum statistics} of excitations.

\textit{\underline{Digression}:} Let $W$ and $W'$ be two Wilson networks both creating excitations 
describing $n-k$ electrons and $k$ holes ($k=0,1, \dots,n$) located at \textit{disjoint} points inside 
some region $\mathcal{O}\subset \Omega_0$, with $q_{\mathcal{O}}(W)=q_{\mathcal{O}}(W')$. 
Let $\widetilde{W}$ be an \textit{arbitrary} Wilson network with the property that the networks
$(W\cdot W') \circ \Theta(\widetilde{W})$ and $\mathcal{B}_{\mathcal{O}}(W \cdot W') \circ \Theta(\widetilde{W})$ 
are \textit{gauge-invariant}, where $\mathcal{B}_{\mathcal{O}}(W \cdot W')$ arises from 
$W\cdot W'$ by braiding all lines of the two networks with endpoints inside $\mathcal{O}$, 
and \textit{only} those, in such a way that the endpoints of all lines of $W$ ending
inside $\mathcal{O}$ are exchanged with the endpoints of all lines of $W'$ ending in $\mathcal{O}$, but 
\textit{without any lines twisted around or crossing each other}. Then \eqref{4.15} and \eqref{4.17} imply that
$$\langle \widetilde{W} \Vert \mathcal{B}_{\mathcal{O}}(W\cdot W') \rangle = \text{exp}(i\pi n^{2}) \langle 
\widetilde{W}\Vert (W\cdot W') \rangle = (-1)^{q_{\mathcal{O}}} \langle \widetilde{W}\Vert (W\cdot W') \rangle\,.$$
This is the standard \textit{connection between electric charge and statistics} of excitations in a 2D incompressible
electron gas.\quad {\large{$\square$}}

Consider a Wilson network $W$ with support in $\Lambda_{-}$ that has just a single line, $\gamma_{p}$, 
ending in a point $p$ contained in a region $\mathcal{O} \subset \Omega_0$, and let $q:=(q^{\alpha})_{\alpha=1}^{N}$ 
denote the quantum numbers (fluxes) dual to the potentials $B_{\alpha}$ attached to the line $\gamma_{p}$. 
Then $W$ is given by the Wilson line
$$\text{exp}\big(i \sum_{\alpha=1}^{N} \sqrt{2 \pi} q^{\alpha}\int_{\gamma_{p}} B_{\alpha}\big)\,$$
multiplied by lines or flux tubes contained in $\Lambda_{-}$ that do \textit{not} intersect $\Omega_0$ 
in the region $\mathcal{O}$. It follows from Eq.~\eqref{4.18} that 
\begin{equation}\label{4.19}
q_{\mathcal{O}}(W)= \sum_{\alpha=1}^{N} Q_{\alpha} \,q^{\alpha} = Q\cdot q\,.
\end{equation}
It is easy to argue that \textit{quantum numbers}, $q=(q^{\alpha})_{\alpha=1}^{N}$, \textit{corresponding to 
multi-electron/hole excitations,} form a module, $\Gamma$, over $\mathbb{Z}$ of rank $N$, i.e., 
\textit{an integral lattice of rank} $N$, which we call a \textit{``Hall lattice''}; (to see this, consider the additivity 
of the electric charge and other ``pseudo-charge'' quantum numbers under composition 
of excitations). By \eqref{4.19}, the ``vector'' $Q= (Q_1,...,Q_N)$ must be an integer-valued, $\mathbb{Z}$-linear 
functional on $\Gamma$, i.e., an element of the dual lattice, $\Gamma^{*}$. 

The lattice $\Gamma$ is equipped with an odd-integral quadratic form, 
$$\langle q^{(1)}, q^{(2)}\rangle:= \sum_{\alpha} q^{(1) \alpha}\cdot q^{(2) \alpha}\,, \quad q^{(1)}, q^{(2)} \in \Gamma\,.$$
That this defines an odd-integral quadratic form is shown as follows: 
Braiding two Wilson lines with quantum numbers $q^{(1)}=q^{(2)}=q \in \Gamma$ ending in $\Omega_0$
yields a phase factor $\text{exp}\big(i\pi \langle q, q \rangle\big)$, which must be $=1$ if $Q\cdot q$ is even, 
and $=-1$ if $Q\cdot q$ is odd (connection between electric charge and quantum statistics).

If $q$ is the vector of quantum numbers corresponding to a single electron/hole then
\begin{equation}\label{4.20}
Q\cdot q = \mp 1, \text{    and   }\,\, \text{exp}\big(i \pi \langle q,q \rangle \big) = -1\,.
\end{equation}
It follows that $\langle \cdot, \cdot\rangle$ is odd-integral, and $Q$ is a ``visible'' vector in $\Gamma^{*}$.
Since $Q\in \Gamma^{*}$ and $\Gamma$ is an (odd-) integral lattice, we conclude that (re-instating $e^{2}/h$)
\begin{equation}\label{4.21}
\frac{h}{e^{2}} \,\sigma_{H} = Q\cdot Q \equiv \sum_{\alpha=1}^{N} Q_{\alpha}^{2} \,\,\,\text{ is a } rational \text{ number}\,.
\end{equation}
\begin{center}
\includegraphics[width=16cm]{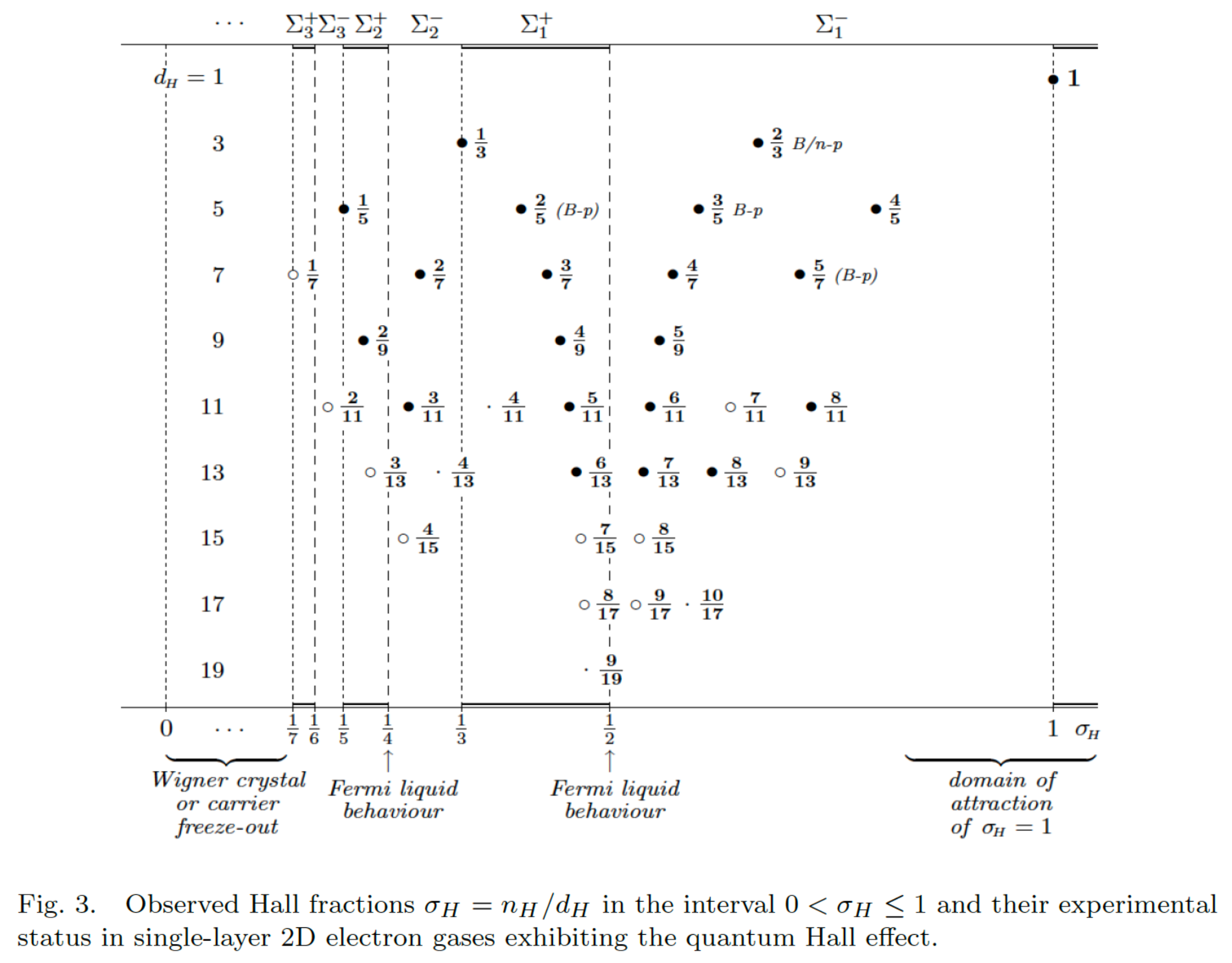}
\end{center}
Our task is then to classify the pairs $(\Gamma, Q \in \Gamma^{*})$, using \textit{invariants} of these data that encode
physical properties of the corresponding incompressible 2D electron liquid. This (quite non-trivial) task has been carried 
out in collaboration with U.~M.~Studer and E.~Thiran, during the period from 1992 till 1994; see 
\cite{F-Thiran, Les Houches 94, ICM 94}). I will not go into any details here. Among the results of our efforts
we have predicted the table of Hall fractions in the interval $(0,1]$ shown in Figure 3; fractions indicated by a bullet (full dot) 
have been experimentally observed in incompressible electron liquids. In Figure 3, the Hall conductivity, 
$\sigma_H = \frac{e^{2}}{h}\cdot \frac{n_H}{d_H}$, is plotted horizontally ($n_H$ and $d_H$ are co-prime integers), and 
$d_H$ is plotted vertically; units are chosen such that $e^{2}/h =1$.\\

\noindent \textbf{Comments.}
The Hall lattices $\Gamma$ corresponding to fractions observed in the intervals $\Sigma^{\pm}_{\ell}, \ell=2,3, \dots,$ 
(see Figure 3) are related to those corresponding to fractions in the intervals $\Sigma^{\pm}_1$ by the so-called 
\textit{shift map}; see \cite{F-Thiran, ICM 94}. 
The pattern of Hall fractions in $\Sigma^{-}_{\ell}$ is much richer than the pattern in $\Sigma^{+}_{\ell}$. 
The Hall lattices $\Gamma$ corresponding to the Hall fractions $\frac{1}{3}, \frac{2}{5}, \frac{3}{7}, \dots$ 
contain the $A$ root lattices as their Witt sublattices\footnote{See, e.g., \cite{Conway-Sloane} for information
about integral lattices and their invariants.} and are \textit{unique}. Among the Witt sublattices of the Hall
lattices corresponding to Hall fractions in $\Sigma^{-}_1$ one also finds the $D$- and the $E$- root lattices.
Some of the fractions (e.g., $\frac{2}{3}$) correspond to two or more \textit{distinct} pairs $(\Gamma, Q)$. The
experimental manifestation of this ambiguity is that, at a fixed value of the filling factor, transitions between 
\textit{distinct} Hall insulator states with the \textit{same} value of the Hall conductivity are observed 
when the external magnetic field is tilted (which affects the orientation of the spins of the electrons).

The story just told can also be presented from the point of view of edge degrees of freedom.
The chiral anomaly \eqref{Hall cond} and expression \eqref{edge=bulk} imply that there must exist several ($N$) species 
of gapless chiral quasi-particles 
propagating along the edge. They are described by $N$ chiral scalar Bose fields 
$\lbrace \varphi^{\alpha} \rbrace_{\alpha=1}^{N}$ with propagation speeds $\lbrace v_{\alpha} \rbrace_{\alpha=1}^{N}$, 
with the properties that
\begin{enumerate}
\item{the chiral electric edge current operator and the Hall conductivity are given by the formulae
$$
\frak{j}^{\mu}_{edge}= e \sum_{\alpha=1}^{N}\frac{Q_{\alpha}}{\sqrt{2 \pi}} \, \partial^{\mu}\varphi^{\alpha}, \quad 
 \sigma_{H}^{edge}=\frac{e^{2}}{h}Q\cdot Q^{T}\overset{!}{=} \sigma_H\,, \,\,\text{ with }\,\,Q:=(Q_1,\dots,Q_{N}) ;
$$
and}
\item{multi-electron/hole states localized at an edge are created by the vertex operators
\begin{equation}\label{4.21}
:\text{exp }i\left(\sum_{\alpha=1}^{N} \sqrt{2\pi}\, q_{\alpha} \varphi^{\alpha}\right):\,\, \text{ where  } \,\, \,
q= \begin{pmatrix}q_{1}\\ \vdots\\ q_{N} \end{pmatrix} \in \Gamma\,.
\end{equation}
The connection between electric charge and (quantum) statistics valid in 1D chiral ``Luttinger liquids''
implies that $\Gamma$ must be an odd-integral lattice of rank $N$ (which is seen by arguments identical
to the ones described above for the bulk of the electron liquid). If the Witt sublattice of a Hall lattice 
$\Gamma$ contains the root lattice of a Lie algebra $\mathfrak{g}$ then $\mathfrak{g}$ must belong 
to the $A-$, $D-$ or $E$ series; and one finds that the the chiral Luttinger liquid describing the edge 
degrees of freedom gives rise to a Kac-Moody algebra $\widehat{\mathfrak{g}}_{k=1}$ at level 1.

A large family of Hall insulators can be classified by the data encoded into an odd-integral lattice 
$\Gamma$ and a visible vector $Q$ in the dual lattice $\Gamma^{*}$. As already mentioned, 
this implies that the Hall conductivity of such systems has the property that $\frac{h}{e^{2}} \sigma_{H} \in \mathbb{Q}$.
}
\end{enumerate}
\newpage
\noindent \textbf{Remarks.}
\begin{enumerate}
\item[I.]{It is expected that vectors $q^{*}$ belonging to the dual lattice $\Gamma^{*} (\supset \Gamma)$ 
are the quantum numbers of quasi-particles with \textit{fractional charge} and \textit{fractional statistics.}}
\item[II.]{The electric charges of quasi-particles propagating along the edges/boundaries of a 2D incompressible 
electron liquid can and have been measured by using Mach-Zehnder interferometry (see \cite{Glattli}, and
\cite{FLS, BF} for theory).
}
\end{enumerate}
Further details can be found in \cite{F-Thiran, Les Houches 94, ICM 94}. For a classification of \textit{``non-abelian''} 
Hall insulators, see \cite{FPSW}. Such insulators are expected to exhibit quasi-particles with 
\textit{non-abelian braid statistics} and may therefore be of interest for purposes of topological quantum computing.

\section{ Induced Chern-Simons actions, dualities, and 2D chiral photonic wave guides}
We consider a model of 3D quantum electrodynamics ($QED_3$), namely a relativistic quantum field theory 
of an odd number of Dirac fermions of electric charge $e$ described by 2-component spinor fields, $\psi_{\alpha}$, 
of masses $M_{\alpha} \not= 0,\, \alpha =1,2,...,2n+1$, propagating in a three-dimensional space-time, 
$\Lambda \,(= \Omega\times \mathbb{R})$, and minimally coupled to an electromagnetic vector potential $A$. 
This theory breaks the symmetries of time reversal and reflection in lines. Integrating over the 
degrees of freedom of the Dirac fermions, we find that the effective action of the vector potential $A$ 
is given by
\begin{align}\label{5.1}
S_{\Lambda}(A)= &\sum_{\alpha=1}^{2n+1} \ell\text{n} [\text{det}_{ren}\big((\partial_{\mu} - eA_{\mu})\gamma^{\mu} +M_{\alpha}\big)]\nonumber \\
=& \sum_{\alpha=1}^{2n+1} Tr\,\ell\text{n}\big({\bf{1}} - G_{M_{\alpha}}\,eA_{\mu}\gamma^{\mu}\big)\,,
\end{align}
where the matrices $\gamma^{\mu}, \mu=0,1,2,$ are $2\times 2$ Dirac matrices, and $G_{M}$ is the propagator of a 
free 2-component Dirac spinor with mass $M\not= 0$ propagating in $\Lambda$; (constants are ignored on the right 
side of \eqref{5.1}). One may expand the logarithm on the right side of \eqref{5.1} in powers of $A$.
For a large value of the mass $M$, the leading term in $Tr \ell\text{n}\big({\bf{1}}+G_{M} A_{\mu}\gamma^{\mu}\big)$ 
is the one quadratic in $A$, which can be calculated explicitly without much difficulty.\footnote{This has been done
in unpublished work on $QED_3$, by J. Magnen, the late R. S\'{e}n\'{e}or and myself, in 1976. Explicit expressions 
were first published by S.~Deser, R.~Jackiw and S.~Templeton in \cite{DJT}; see also \cite{Redlich}.}
It is given by 
\begin{equation}\label{5.2}
sgn(M)\,\frac{e^{2}}{8\pi \hbar} \int_{\Lambda} A\wedge dA + \text{boundary term},
\end{equation}
i.e., by a Chern-Simons term corresponding to a Hall conductivity $\sigma_{H}=\pm \frac{1}{2}\cdot \frac{e^{2}}{h}$. Terms of higher order in $A$ tend to $0$, as $M\rightarrow \infty$.
I will not reproduce the calculations leading to \eqref{5.2}; but see \cite{DJT, Redlich}.

If the electromagnetic field is treated as a dynamical quantum field one must add a Maxwell 
term to the induced Chern-Simons action appearing on the right side of \eqref{5.1}. Thus the 
total effective action functional of $QED_3$ is given by
\begin{align}\label{5.3}
S_{\Lambda}(A)= &\sum_{\alpha=1}^{2n+1}sgn(M_{\alpha}) \big\{ \frac{e^{2}}{8\pi \hbar} \int_{\Lambda} A\wedge dA + \Gamma_{\partial \Lambda}(A_{\Vert})\big\} + \nonumber \\
+ &\int_{\Lambda} \big[\varepsilon \underline{E}^{2} - \mu^{-1} B^{2}\big]d^{3}x + \text{less relevant terms}\,.
\end{align}
Neglecting terms of higher than quadratic order in the electromagnetic vector potential $A$ on the right side
of \eqref{5.3}, $S_{\Lambda}(A)$ is a quadratic form in the vector potential $A$. If $\Omega$ is a cube and
periodic boundary conditions are imposed at $\partial \Omega$ (i.e., $\Omega$ is a torus) then this quadratic 
form can be diagonalized by Fourier transformation (exercise). This yields an explicit (momentum-space) 
expression for the 2-point functions of the electromagnetic field tensor; general $n$-point functions can 
be expressed as sums of products of 2-point functions by appealing to Wick's theorem. The denominators 
of the imaginary-time (euclidian) 2-point functions of the components, $F_{\mu \nu}$, of the electromagnetic 
field tensor in momentum space are given by $[k^{2}+\text{const.}e^{2}]$, where $k$ is the momentum variable. 
This shows that, in this model of $QED_3$, \textit{photons} have a \textit{strictly positive mass} proportional to $e$.

If space-time $\Lambda$ has a boundary then the effective action of the electromagnetic field has a boundary 
term given by the anomalous chiral action $\Gamma_{\partial \Lambda}(A_{\Vert})$ cancelling the anomaly 
of the Chern-Simons term in \eqref{5.3}, as discussed in \eqref{boundary action} and \eqref{4.16}.

It is argued that, in certain planar systems of condensed matter, there exist quasi-particles with low-energy 
properties mimicking those of 2-component Dirac fermions. An example is ``doped'' graphene (see, e.g., 
\cite{graphene}, and references given there). The low-energy properties of such systems can be 
described by a theory approaching the model of $QED_{3}$ introduced above in the infrared domain.

In planar systems (i.e., systems in three space-time dimensions), the electromagnetic vector potential $A$ 
and the vector potential, $B$, of the conserved electric current density, $J =\nabla \wedge B$, play
\textit{dual roles}: Under the replacements  
$$A \mapsto B, \quad B\mapsto A,$$
 \textit{conventional time-reversal invariant 2D insulators} are mapped to \textit{2D superconductors}, 
 and \textit{electronic Hall- or Chern insulators} are mapped to \textit{gapped photonic wave guides} exhibiting 
 extended chiral electromagnetic surface waves; and \textit{conversely} (see \cite{Les Houches 94}).
 
As an example we consider the duality between certain Hall- or Chern insulators and gapped photonic wave guides.
We define
 \begin{equation}\label{5.4}
\widetilde{S}_{\Lambda}(B):= \frac{1}{2\sigma_{H}} \int_{\Lambda} B \wedge dB + \text{ boundary action }\, +\, \text{ less relevant terms}, \,\,\,\,\sigma_H:= \frac{e^{2}}{4\pi \hbar}\,.
\end{equation}
Then we have the duality, valid in the infrared, between $QED_{3}$ \textit{with induced Chern-Simons action} and 
\textit{the quantum theory of currents in some Hall insulator}.
This is elucidated by functional Fourier transformation:
\begin{equation}\label{5.5}
e^{i\,S_{\Lambda}(A)}= \mathcal{N}^{-1} \int e^{i\widetilde{S}_{\Lambda}(B)}\,e^{i\int_{\Lambda} A\wedge dB} \mathcal{D}B\,,
\end{equation}
where $\mathcal{N}$ is a normalization factor, and conversely. We may view the current driven through a wave guide 
with broken time-reversal invariance as a \textit{``classical control variable''}, while the electromagnetic field is treated 
as quantized \textit{dynamical} degrees of freedom. The response equations of such a wave guide are given by
\begin{align}\label{5.6}
\langle F_{\mu\nu}(x) \rangle_{B}= \varepsilon_{\mu\nu\lambda} \frac{\delta \widetilde{S}_{\Lambda}(B)}{\delta B_{\lambda}(x)}
=\sigma_{H}^{-1} \varepsilon_{\mu\nu\lambda}\, j^{\lambda}(x)\,.
\end{align}
The boundary action on the right side of Eq.~\eqref{5.4} is 
given by $ \frac{1}{\sigma_{H}} \Gamma_{\partial \Lambda}^{(\pm)}(B\vert_{\partial \Lambda}),$
with
$$\Gamma_{\partial \Lambda}^{(\pm)}(b) :=  \frac{1}{2} \int_{\partial \Lambda} [b_{+}b_{-} - 
2b_{\pm} \frac{\partial^{2}_{\pm}}{\Box} b_{\pm}] du^{+}\,du^{-}$$
in light-cone coordinates $(u^{+}, u^{-})$ on $\partial \Lambda$, with $B\vert_{\partial \Lambda}= b_{+}du^{+} + b_{-}du^{-}$.
The sign of $\sigma_H$ and the choice of $\pm$ in $\Gamma_{\partial \Lambda}^{(\pm)}$ depend 
on the chirality of the electromagnetic edge waves.
The boundary action $\Gamma_{\partial \Lambda}^{(\pm)}(b)$ is the generating functional of Green functions of \textit{gapless edge modes of the quantized electromagnetic field}
propagating chirally around the boundary of the wave guide. In the bulk of such wave guides, the electromagnetic field
is gapped, as remarked above.

It would be tempting to continue with a discussion of various further topics, such as the \textit{theory of rotating Bose gases} 
in two dimensions and their Hall effects. This topic started with my work with U.~M.~Studer and E.~Thiran 
(see \cite{Les Houches 94}). Much further work on 2D Bose gases was carried out in \cite{Bose gases} (and references
given there). The role of \textit{gravitational anomalies} in an analysis of heat transport in 2D systems of condensed 
matter physics has been elucidated in \cite{K-W}.

Five-dimensional QED, a close cousin of the theory introduced in \eqref{5.1} through \eqref{5.3}, will 
make a brief appearance in Section 8 and might be of interest, for example, in cosmology.

\section{Chiral Spin Currents in Planar Topological Insulators}
So far, we have ignored electron spin, in spite of the fact that there are 2D gapped electron liquids exhibiting the 
fractional quantum Hall effect where electron spin plays an important role, as mentioned in subsection 4.2. (We won't 
study these systems any further; but see \cite{FS-RMP, F-Thiran, Les Houches 94}, and references given there.)

In this section we consider \textbf{time-reversal invariant 2D topological insulators} (2D TI) exhibiting chiral spin 
currents.\footnote{A general reference where the role of electron spin and spin currents and the spin Hall 
effect are analyzed is \cite{Dyakonov}.} -- We start from the \textit{Pauli equation} for the wave function, $\Psi_t$, 
of a spinning electron
\begin{equation}\label{6.1}
i\hbar D_{0}\Psi_{t}=-\frac{\hbar^{2}}{2m} g^{-1/2}D_{k}\,\big(g^{1/2}g^{kl} \, D_{l} \Psi_{t}\big)\,,
\end{equation}
where $m$ is the effective mass of an electron, $(g_{kl})$ is the Riemannian metric on sample space, 
$\Omega$, an orthonormal frame bundle is introduced on space-time $\Omega \times \mathbb{R}$ enabling 
one to define \mbox{2-component} \textit{Pauli spinors}, $\Psi$, and, in particular, to introduce notions of spin-up, 
$\uparrow$, and \mbox{spin-down, $\downarrow$:}
$$\Psi_{t}(x)= \begin{pmatrix} \psi^{\uparrow}_{t}(x)\\
\psi^{\downarrow}_{t}(x) \end{pmatrix} \in L^{2}(\Omega, d\,vol.)\otimes \mathbb{C}^{2}\,.$$
Furthermore, the symbols $D_{\mu},\, \mu=0,\dots, d, d=2 \text{ or }3,$ are covariant derivatives
\begin{equation}\label{6.2}
i\hbar D_{0}= i\hbar \partial_{t} + e \varphi - \vec{W}_{0}\cdot \vec{\sigma}\,,\quad \vec{W}_{0}= \mu c^{2} \vec{B}+ \cdots,
\end{equation}
where $\varphi$ is the electrostatic potential acting on the electrons, $\vec{W}_{0}\cdot \vec{\sigma}$ is the Zeeman 
coupling of the spin of electrons to the zero-component, $\vec{W}_0$, of an $SU(2)$-gauge field; $\vec{W}_0$ 
describes an external magnetic field and/or a Weiss exchange field \cite{AFFS}. Furthermore,
\begin{equation}\label{6.3}
\frac{\hbar}{i}D_{k}= \frac{\hbar}{i}\nabla_{k} + eA_{k} - \vec{W}_{k}\cdot \vec{\sigma}+ \cdots\,,
\end{equation}
where $\vec{A}$ is the electromagnetic vector potential, and the dots stand for further terms 
present in a moving background (which we ignore in the following, but see \cite{Les Houches 94}); 
moreover, $\vec{W}_k$ is the k-component of an $SU(2)$-gauge field, which, for the systems considered 
in this section, describes spin-orbit interactions, hence is given by
\begin{equation}\label{6.4}
\vec{W}_{k}\cdot \vec{\sigma}:= \big[(-\tilde{\mu} \vec{E}+ \cdots)\wedge \vec{\sigma}\big]_{k},
\end{equation}
where $\tilde{\mu}= \mu + \frac{e\hbar}{4mc^{2}}$\,; (the term $\frac{e\hbar}{4mc^{2}}$ is due to Thomas precession).

We observe that the Pauli equation \eqref{6.1} displays perfect {$U(1)_{em}\times SU(2)_{spin}$ - gauge invariance},
$A=A_0 dt + A_1 dx^{1}+ \dots + A_d dx^{d}$ is the $U(1)_{em}$- connection, while 
$\vec{W}= \vec{W}_0 dt + \vec{W}_1 dx^{1} + \dots \vec{W}_d dx^{d}$ is the $SU(2)_{spin}$- connection,  
with $x=(t\equiv x^{0}, x^{1}, \dots, x^{d}), d= 2  \text { or }3,$ the space-time coordinates on 
$\Lambda= \Omega \times \mathbb{R}$.\footnote{See \cite{FS-RMP, Les Houches 94}
for a more detailed discussion.}

In the following, we consider an interacting electron gas confined to a planar region $\Omega$, 
setting $d=2$ and  $x^{1}\equiv x, x^{2}\equiv y$. We assume that $\vec{B} \perp \Omega$ and $\vec{E} \Vert \Omega$. 
Then the $SU(2)_{spin}$ - connection, $\vec{W}_{\mu}$, is given by
\begin{equation}\label{6.5}
W^{3}_{\mu}\cdot \sigma_{3}, \quad \text{with  } W^{K}_{\mu}\equiv 0, \text{  for   } K=1,2\,,
\end{equation}
with $W^{3}_{0}$ given by \eqref{6.2} and $W^{3}_k, k=1,2,$ given by \eqref{6.4}.
From \eqref{6.5} we conclude that, for the systems considered here, parallel transport of Pauli spinors splits 
into parallel transport for the spin-up ($\uparrow$) and the spin-down ($\downarrow$) component. 
The spin-up component, $\psi^{\uparrow}$, of a Pauli spinor $\Psi$ couples to the \textit{abelian} 
connection $a+w$, while $\psi^{\downarrow}$ couples to $a-w$, where 
\begin{equation}\label{6.5-1}
a_{\mu}=-eA_{\mu}, \, \text{ and  }\, w_{\mu}= W^{3}_{\mu}\,,
\end{equation}
see \eqref{6.2} and \eqref{6.4}. Under time reversal, $T$,
\begin{equation}\label{6.6} 
a_{0} \rightarrow a_{0}, \,a_{k}\rightarrow -a_{k}, \,\text{  but  }\, w_{0}\rightarrow -w_{0}, \,w_{k} \rightarrow w_{k}\,,
\end{equation}
as is well known. The dominant term in the effective action of a 2D time-reversal invariant topological insulator, 
with $\vec{W}$ as in \eqref{6.5}, is given by a combination of Chern-Simons terms. If either $w\equiv 0$ or 
$a\equiv 0$ a Chern-Simons term in $a$ or in $w$ \textit{alone} would \textit{not} be time-reversal invariant. 
Hence, assuming time-reversal invariance and $w \equiv 0$, the dominant term would be given by
\begin{equation}\label{6.7}
S_{\Lambda}(A)= \int_{\Lambda} dt\, d^{2}x \lbrace \varepsilon \underline{E}^{2} - \mu^{-1}B^{2} \rbrace,
\end{equation}
which is the effective action of a \textit{conventional insulator}.

But, in the presence of both connections, $a$ and $w$, a combination of \textit{two} Chern-Simons actions 
\textbf{is} time-reversal invariant. In this case the leading (``most relevant'') contribution to the (bulk) effective action
is given by
\begin{align}\label{6.8}
S_{\Lambda}(a,w) =& \frac{\sigma}{2} \int_{\Lambda} \lbrace (a+w) \wedge d(a+w) - (a-w) \wedge d(a-w) \rbrace 
\nonumber \\
=& \sigma \int_{\Lambda} \lbrace a\wedge dw + w \wedge da \rbrace \,,
\end{align}
where we have neglected the boundary (edge) actions.\footnote{Note that if $\vec{W}$ is expressed in terms of 
$\vec{B}$ and $\vec{E}$, as in \eqref{6.2}, \eqref{6.4}, \eqref{6.5}, one recovers expression \eqref{6.7}. 
The effective action in \eqref{6.8} first appeared in a paper with U.~M.~Studer \cite{FS-RMP} in 1993\,!} 
The gauge fields $a$ and $w$ transform \textit{independently} under gauge transformations preserving condition \eqref{6.5}, and the action in \eqref{6.8} is \textit{anomalous} under these gauge transformations, for a 2D sample space
$\Omega$ with non-empty boundary. We have learned in Section 4 that the anomalous boundary action,
\begin{equation}\label{6.9}
\sigma[\Gamma_{\partial \Omega \times \mathbb{R}}^{+} \big( (a + w)_{\Vert}\big) - \Gamma_{\partial \Omega \times \mathbb{R}}^{-}\big( (a - w)_{\Vert}\big)]\,,
\end{equation}
cancels the anomalies of the bulk action; see Eqs.~\eqref{boundary anomaly} and \eqref{boundary action}. 
This boundary action is the generating functional of connected Green functions of \textit{two counter-propagating 
anomalous chiral edge currents}.
One of the two counter-propagating edge currents is carried by electrons with spin up ($\uparrow$), i.e., polarization in
the $+3$-direction $\perp \Omega$, the other one is carried by electrons with spin down ($\downarrow$). These edge 
currents are exchanged by time reversal. Thus, a net \textit{chiral spin current}, $s^{3}_{edge}$, can be excited to 
propagate along the edge of such a system. 
The bulk response equations (analogous to the Hall-Chern-Simons law \eqref{general Hall}) are given by
\begin{equation}\label{6.10}
j^{k}(x)= 2\sigma \varepsilon^{k \ell} \partial_{\ell} B(x), \,\,\,\,s^{\mu}_{3}(x)=\frac{\delta S_{\Lambda}(a,w)}{\delta w_{\mu}(x)} = 2\sigma \varepsilon^{\mu\nu\lambda} F_{\nu \lambda}(x)\,,
\end{equation}
where we have expressed the $SU(2)$-gauge field $\vec{W}$ in terms of the magnetic induction $\vec{B}=(0,0,B_3)$
and the electric field $\vec{E}=(E_1, E_2, 0)$. As in Section 4, Eqs.~\eqref{puzzle}, \eqref{edge currents} and 
\eqref{2D anomaly}, these equations could again be used to deduce the existence of the edge spin-currents
discovered above.

We should ask what kinds of quasi-particles in the bulk of planar materials could produce the bulk Chern-Simons 
terms in \eqref{6.8}. Given our findings in Section 5, it is tempting to argue that a 2D time-reversal invariant 
topological insulator with a bulk effective action as given in \eqref{6.8} must exhibit two species of charged 
quasi-particles in the bulk, with one species (spin-up, $\uparrow$) related to the other one (spin-down, $\downarrow$) 
by time-reversal, and each species has two degenerate states per wave vector mimicking a two-component Dirac 
fermion at small energies, leading to the quantization \mbox{of $\sigma$.}

Materials of this kind have been produced and studied in the laboratory of L. Molenkamp in W\"{u}rzburg (see 
\cite{Molenkamp}). The experimental data are not very clean, the likely reason being that, due to 
magnetic impurities in the material and/or electric fields in the direction $\perp \Omega$, the condition in \eqref{6.5} 
is violated, i.e., the $SU(2)$-gauge field $\vec{W}_{\mu}$ has non-vanishing components $W^{K}_{\mu}$ and is 
genuinely \textit{non-abelian}. In this situation, the spin current is \textit{not} conserved, anymore, 
(but continues to be covariantly conserved, see \cite{FS-RMP}), and time reversal invariance is broken.\\

The approach to 2D time-reversal invariant topological insulators sketched in this section can be generalized 
as follows. We consider a state of matter exhibiting a bulk-spectrum of two species of quasi-particles related to 
one another by time-reversal. In order to analyze the transport properties of such a state, one should study its 
response when one species is coupled to a (real or virtual, abelian or \textit{non-abelian}) external gauge 
field\footnote{often dubbed ``Berry connection''} $W^{+}$ and the other one to a gauge field $W^{-}$ related 
to each other by time-reversal according to
$$(W_{0}^{+})^T = W_{0}^{-}, \quad (W_{k}^{+})^{T} = - W_{k}^{-}.$$
If the leading term in the effective action for the gauge fields $W^{+}$ and $W^{-}$ is given by the sum of two identical 
Chern-Simons terms, but with \textit{opposite} signs, time-reversal invariance is manifestly preserved, and one 
concludes that there are \textit{two counter-propagating chiral edge currents} generating current (Kac-Moody) 
algebras\footnote{at level 1, for non-interacting electrons} based on a Lie group given by the gauge group 
of the gauge fields $W^{\pm}$. For non-interacting electrons, this group can be determined from band theory.

If one gives up the requirement of time-reversal invariance one arrives at a general theory of \textit{chiral states} 
in 2D (planar) systems of condensed matter physics. If $\vec{W}$ is an $SU(2)$-gauge field coupling to the spin 
of electrons, as in \eqref{6.2} and \eqref{6.4}, one obtains a framework well suited to describe \textit{chiral spin liquids} 
(see \cite{Les Houches 94} and references given there).

\section{Anomalous current commutators and the chiral magnetic effect}
I recall the definition of left-handed and right-handed currents and the anomalous current
commutators described in Section 2 for theories in four space-time dimensions, namely
\begin{align}\label{7.1}
J^{\mu}_{\ell}:= \frac{1}{2}\big[J^{\mu}+&J^{\mu}_{5}\big], \quad J^{\mu}_r:= \frac{1}{2}\big[J^{\mu}-J^{\mu}_{5}\big]
\nonumber\\
\big[J^{0}_{\ell/r}(\vec{y},t), J^{0}(\vec{x},t)\big] =& \pm i\frac{\alpha}{8\pi^{2}}\big(\vec{B}(\vec{y},t)\cdot \vec{\nabla}_{\vec{y}}\big)\,
\delta(\vec{x}-\vec{y})\,;
\end{align}
see \cite{CAA} and references given there, and \cite{F-T} for a derivation using functional
integrals.

I proceed to recall the \textit{chiral magnetic effect} and its derivation from the anomalous commutator stated in \eqref{7.1}.
For this purpose I consider a theory of charged, massless Dirac fermions in four space-time dimensions.
This theory has a conserved vector current, $J^{\mu}$, satisfying the continuity equation
$$\partial_{\mu} J^{\mu} =0\,.$$
This equation implies that there exist two vector fields, $\vec{\varphi }(x)$ and $\vec{\chi}(x)$, such that
\begin{equation}\label{8.1}
J^{0}(x) = \frac{e}{2\pi} \vec{\nabla}\cdot \vec{\varphi}(x), \quad \vec{J}(x)=
 - \frac{e}{2\pi} \Big\{\frac{\partial}{\partial t}\vec{\varphi}(x) + \vec{\nabla} \times \vec{\chi}(x)\Big\}\,,
\end{equation}
with $e$ the electric charge of the fermions.
If $H$ denotes the Hamiltonian of the system then the Heisenberg equations formally imply that
\begin{equation}\label{8.2}
\frac{\partial}{\partial t} \vec{\varphi}(x) = \frac{i}{\hbar} [H, \vec{\varphi}(x)] \,.
\end{equation}
We define chiral charges
\begin{equation}\label{8.3}
N_{\ell/r}:= \int d^{3}x\, \widehat{J}^{0}_{\ell/r}(t, \vec{x}), 
\end{equation}
where
$$
\widehat{J}^{\mu}_{\ell/r}:= J^{\mu}_{\ell/r} \mp \frac{e^{2}}{16 \pi^{2}} \varepsilon^{\mu\nu\rho\lambda}A_{\nu}F_{\rho\lambda}\,.
$$
The second term on the right side of this equation is dual to the Chern-Simons 3-form, $A\wedge dA$, that we are already 
familiar with. Since the fermions are assumed to be massless, these charges are \textit{conserved}, as follows 
from Equation \eqref{7} in Section 2, and \textit{gauge-invariant}. 

Henceforth we suppose that all external fields, in particular the electromagnetic field, are \textit{time-independent}.
We may then consider an equilibrium state, $\langle (\cdot) \rangle_{\beta, \underline{\mu}}$, of the system with 
non-vanishing \textit{chemical potentials}, $\underline{\mu}:= (\mu_{\ell}, \mu_{r})$, conjugate to the charges 
$N_{\ell}$ and $N_{r}$, respectively. Our aim is to calculate $\vec{j}(x)= \langle \vec{J}(x) \rangle_{\beta, \underline{\mu}}$, 
using arguments reminiscent of those in Section 3. By equations \eqref{8.1} and \eqref{8.2},
 \begin{equation}\label{8.4}
 \vec{j}(x)= -\frac{ie}{h} \langle [H, \vec{\varphi}(x)]\rangle_{\beta,\underline{\mu}} 
 -\frac{e}{2\pi} \vec{\nabla} \times \langle \vec{\chi}(x) \rangle_{\beta,\underline{\mu}}\,
 \end{equation}
 Formally, the term proportional to $\langle [H, \vec{\varphi}(x)]\rangle_{\beta,\underline{\mu}}$ vanishes, 
 because the equilibrium state is time-translation invariant. However, the field 
 $\vec{\varphi}$ turns out to have ill-defined zero-modes, so that we cannot use the identity 
 $[H, \vec{\varphi}(x)]= H\vec{\varphi}(x)- \vec{\varphi}(x)H$. We must regularize the right side of \eqref{8.4} 
 by giving the field $\vec{\varphi}$ a small mass and then use that
 \begin{equation}\label{8.5}
 \frac{\partial}{\partial t} \vec{\varphi}(x)= \frac{i}{h} \big[H- \mu_{\ell} N_{\ell} - \mu_{r}N_{r} , \vec{\varphi}(x)\big]
 + \frac{i}{h} \big[\mu_{\ell} N_{\ell} + \mu_{r}N_{r}, \vec{\varphi}(x) \big] \,
 \end{equation}
 and that\quad
 $\langle \big[H- \mu_{\ell} N_{\ell} - \mu_{r}N_{r}, \vec{\varphi}(x) \big] \rangle_{\beta, \underline{\mu}}=0.$
Combining this identity with \eqref{8.4} and \eqref{8.5}, we find the \textit{``current sum rule''}
\begin{equation}\label{8.6}
\vec{j}(x)= - \frac{ie}{h}\langle \big[\mu_{\ell} N_{\ell} + \mu_{r}N_{r}, \vec{\varphi}(x) \big] \rangle_{\beta, \underline{\mu}}
-\frac{e}{2\pi} \vec{\nabla} \times \langle \vec{\chi}(x) \rangle_{\beta,\underline{\mu}}\,.
\end{equation}
Recalling formula \eqref{7.1} for the anomalous current commutators and \eqref{8.1}, we conclude that
\begin{equation}\label{8.7}
\big[\widehat{J}^{0}_{\ell/r}(\vec{y},t), \vec{\varphi}(\vec{x},t)\big]= \mp i\frac{e}{4\pi} \vec{B}(\vec{y},t) \,
\delta(\vec{x}-\vec{y}) + \vec{\nabla}\wedge \vec{\Pi}_{\ell/r}(\vec{x}, \vec{y};t)
\end{equation}
Using \eqref{8.3}, \eqref{8.6} and \eqref{8.7}, we find that the current density $\vec{j}$ is given by the equation
\begin{equation}\label{8.8}
\vec{j}(x)= \frac{e^{2}}{4\pi h}\big(\mu_{\ell}-\mu_{r}\big) \vec{B}(x) + \vec{\nabla} \times \vec{\frak{B}}(x)\,,
\end{equation}
where the second term, proportional to $\vec{\frak{B}}$, on the right side describes a persistent current not related
to the magnetic induction $\vec{B}$, which we will not consider any further. Equation \eqref{8.8} describes the 
\textit{chiral magnetic effect} first considered in \cite{Vilenkin} (see \cite{ACF} for a derivation along the lines 
of the arguments presented above). A more conceptual treatment of this effect is given in Section 8.

An effect related to the chiral magnetic effect might be observed in systems of spinning particles propagating in a 
\textit{rotating (non-inertially moving) background}, such as leptons in a rapidly rotating star.

\subsection{A possible manifestation of the chiral magnetic effect in cosmology}
In this subsection we sketch ideas concerning the evolution of magnetic fields in an expanding 
Universe, i.e., with a Hubble constant $H\geq 0$, filled with a plasma of charged particles, assuming 
that there exists an asymmetry between right-handed and left-handed fermions, so that the chiral 
chemical potential $\mu_5:=\mu_{\ell}- \mu_{r}$ does not vanish. In the following we use units such 
that the speed of light $c=1$.

Faraday's induction law and the absence of magnetic monopoles are encoded in the homogeneous Maxwell equations
\begin{equation*}
\vec{\nabla}\wedge \vec{E} =- \dot{\vec{B}} -H \vec{B}, \quad \vec{\nabla}\cdot \vec{B}=0\,.
\end{equation*}
The Hubble constant $H$ is assumed to be slowly varying in time.
Coulomb's law and the Amp\`ere-Maxwell law are encoded in the inhomogeneous Maxwell equations 
\begin{equation*}
\vec{\nabla}\cdot \vec{E}=j^{0}, \qquad \vec{\nabla}\wedge \vec{B}= \dot{\vec{E}} + H\vec{E}+ \vec{j}\,,
\end{equation*}
where $j=(j^{0}, \vec{j})$ is the vector current density of charged matter. We assume that the leading contribution
to $\vec{j}$ is due to the \textit{chiral magnetic effect}, besides a conventional contribution described by
\textit{Ohm's law}, i.e.,
\begin{equation}\label{current law}
\vec{j}=\frac{\alpha}{8\pi^{2}} \mu_{5} \vec{B} + \sigma \vec{E}\,,\qquad \alpha = \frac{e^{2}}{\hbar}\,,
\end{equation}
where $\sigma$ is the \textit{Ohmic conductivity} of the plasma.
Taking the \textit{curl} of the Amp\`{e}re-Maxwell law, with $\vec{j}$ as in \eqref{current law}, and
using Faraday's induction law we find that 
\begin{equation}\label{evol B}
\Box\, \vec{B}   +\frac{\alpha}{8\pi^{2}}\mu_{5}\vec{\nabla}\wedge \vec{B}+ (2H+\sigma) \dot{\vec{B}} + 
\mathcal{O}(\dot{H}, \sigma H, H^{2})=0\,,
\end{equation}
where $\Box$ is the d'Alembert operator. From here on, the terms $\mathcal{O}(\dot{H}, \sigma H, H^{2})$ 
in this equation are neglected.
We solve Equation \eqref{evol B} by Fourier transformation, using the ansatz
$$\vec{B}(\vec{x},t)=\vec{b} e^{i(kz- \omega t)}, \,\,\text{with   }\, \vec{b} \perp z\text-{axis}\,,$$
as required by the equation $\vec{\nabla}\cdot \vec{B}=0$. 
Setting $\frac{\alpha}{8\pi^{2}}\mu_{5}=: \tilde{\mu}$, the frequency $\omega(k)$ 
is found to be given by
$$\omega(k)=- i(H+\frac{\sigma}{2})\pm \sqrt{-(H+\frac{\sigma}{2})^{2}+k^{2}\pm \tilde{\mu}\vert k \vert}\,.$$
We observe that the assumptions that $H\geq 0$ (expansion of the Universe) and $\sigma>0$ 
(matter is a conducting plasma) lead to an \textit{exponential damping} of $\vec{B}$ in time, 
\textit{provided} that $\vert k \vert > \tilde{\mu}\,.$
But if $\vert{k} \vert< \tilde{\mu}$ there exists a solution for which $\omega$ has a 
\textit{positive} imaginary part, and the solution of \eqref{evol B} \textit{grows exponentially}. 
This instability might be related to the growth of spatially very slowly varying primordial magnetic fields in the early Universe.
Of course, the value of $\mu_5$ is time-dependent, and it is expected that $\mu_5$ tends to 0, as
time $t$ tends to $\infty$. The instability described here will therefore come to an end, and the
magnetic induction $\vec{B}$ will cease to grow. 

A thorough analysis of this mechanism can be found in \cite{F-Pedrini, BFR, BBFFRS}, and references given there.
 
\section{A five-dimensional cousin of the Hall effect and axion electrodynamics}
We consider a system of charged matter at zero temperature coupled to an external electromagnetic field.
In our derivation of the chiral magnetic effect (see \eqref{8.5} - \eqref{8.8}) we have assumed that the 
external electromagnetic field is \textit{time-independent}. This is usually not the case, and in applcations 
to cosmology and condensed-matter physics, it is unrealistic to assume that $\mu_{5}:=\mu_{\ell}-\mu_{r}$ 
is independent of space and time. It turns out that a \textit{dynamical analogue} of $\mu_5$ is known in particle physics 
under the name of \textit{``axion''}. A very natural way of introducing axions is to study a cousin of the 
\textit{quantum Hall effect} in a five-dimensional (space-time) slab, 
$\Lambda=\Omega\times [0,L]$, with two four-dimensional ``boundary branes,'' $\partial_{\pm}\Lambda\simeq \Omega$
parallel to the $(x^0, x^1, x^2, x^3)$-plane in five-dimensional Minkowski space (where $x^0$ is the time coordinate).
The slab $\Lambda$ is assumed to be filled with \textit{very heavy four-component Dirac fermions} coupled to the 5D 
electromagnetic vector potential, $\widehat{A}$, (viewed as a 1-form on $\Lambda$). Integration over the 
fermion degrees of freedom yields  the effective action for $\widehat{A}$, which has the form
\begin{align}\label{8.9}
S_{\Lambda}(\widehat{A})=& \frac{1}{4L\alpha} \int_{\Lambda} d^{5}x\,\widehat{F}_{MN}(x) \widehat{F}^{MN}(x) + CS_{\Lambda}(\widehat{A}) \nonumber \\
+& \Gamma_{\ell}(\widehat{A}\vert_{\partial_{+}\Lambda})+\Gamma_{r}(\widehat{A}\vert_{\partial_{-}\Lambda})+\dots \,,
\end{align}
where $\alpha$ is dimensionless, $L$ is the width of the 5D slab $\Lambda$, and
\begin{equation}\label{8.10}
CS_{\Lambda}(\widehat{A}):= \frac{\kappa_H}{24 \pi^{2}} \int_{\Lambda} \widehat{A} \wedge \widehat{F} \wedge 
\widehat{F}
\end{equation}
is the \textit{5D Chern-Simons action}.\footnote{In this section we use units such that $\hbar =1$}
This action predicts an analogue of the Hall current given by
\begin{equation}\label{8.10'}
j^{M}= \frac{\kappa_H}{8\pi^{2}}\, \varepsilon^{MNJKL}F_{NJ}\,F_{KL}\,,
\end{equation}
where the dimensionless constant $\kappa_H$ plays the role of the Hall fraction $\frac{h}{e^{2}}\sigma_H$. 
The constant $\kappa_H$, related to the number of species of massive four-component Dirac fermions 
in the bulk, must be an integer for the functional integral to be well defined (assuming  that all fermionic 
degrees of freedom have electric charges that are integer multiples of the elementary electric charge $e$). 
In the following I will set $\kappa_H=1$. Equation \eqref{8.10'} should be compared to equation 
\eqref{general Hall} of Section 4. 

The action functional in equations \eqref{8.9} and \eqref{8.10} describes the electrodynamics of a five-dimensional 
analogue of the quantum Hall effect (5D QHE). Apparently, the 5D QHE can be observed in certain systems of 
condensed-matter physics with virtual dimensions that play the role of space dimensions (see \cite{Zilber} 
and references given there).

In \eqref{8.9}, the functionals $\Gamma_{\ell/r}$ are the anomalous actions of 
left-handed/right-handed Dirac-Weyl fermions located on the boundary branes, $\partial_{\pm}\Lambda$.
The action $\Gamma_{\ell}+\Gamma_r$ on the right side of \eqref{8.9} cancels the anomaly of 
the Chern-Simons action $CS_{\Lambda}(\widehat{A})$; see \cite{CAA}.
 
It is interesting to consider the \textit{dimensional reduction} of the 5D theory with effective action given in
\eqref{8.9} and \eqref{8.10} to a four-dimensional space-time, assuming  that there is a gauge with the property
that $\widehat{A}_M$ is independent of $x^{4}$, for all $M=0,1,2,3.$ We define a field 
$\varphi$ of scaling dimension 0, henceforth called \textit{axion} field, by setting
$$\varphi(x):= \int_{\gamma_x}\widehat{A}\,,$$
where $\gamma_x$ is a path parallel to the $x^{4}$-axis connecting $\partial_{-}\Lambda$ to $\partial_{+} \Lambda$ 
at constant values of $x=(x^{0}, x^{1}, x^{2}, x^{3})$. Then the action functional in \eqref{8.9} becomes
\begin{align}\label{8.11}
S_{\Omega}(A;\varphi)=&\frac{1}{2\alpha} \int_{\Omega} d^{4}x\big[ F_{\mu\nu}(x)F^{\mu\nu}(x) + \frac{1}{L^{2}}\partial_{\mu}\varphi(x)\partial^{\mu}\varphi(x) \big]\nonumber\\
+&\frac{1}{8\pi^{2}}\int_{\Omega} \varphi\,(F\wedge F) + \Gamma_{\Omega}(A)+\dots\,,
\end{align}
Under the present assumptions on the vector potential $\widehat{A}$ 
the term $\Gamma_{\Omega}(A)=\Gamma_{\ell}(A)+\Gamma_{r}(A)$ is \textit{not} 
anomalous and is ignored in the following.

Expression \eqref{8.11} shows that the pseudo-scalar field $\varphi$ can be interpreted as an 
\textit{axion} field, whence its name. One can add a self-interaction term, $U(\varphi)$, to 
the Lagrangian density in \eqref{8.11}, requiring that $U(\varphi)$ be periodic in $\varphi$. 
From \eqref{8.11} we derive the equations of motion for $F_{\mu\nu}= \partial_{\mu}A_{\nu}-\partial_{\nu} A_{\mu}$ 
and $\varphi$.
\begin{equation}\label{axion ED}
\partial_{\mu}F^{\mu\nu}= \frac{\alpha}{4\pi^{2}} \partial_{\mu}\big(\varphi \widetilde{F}^{\mu\nu}\big),\qquad 
L^{-2}\,\Box\,\varphi=\frac{\alpha}{8\pi^{2}} F_{\mu\nu}\widetilde{F}^{\mu\nu}-\frac{\delta U(\varphi)}{\delta \varphi}\,, 
\end{equation}
where $\widetilde{F}^{\mu\nu}$ is the dual field tensor. In terms of the electric and magnetic fields, 
these equations take the form
\begin{align}\label{8.12}
\vec{\nabla}\cdot \vec{B}=&0, \quad \vec{\nabla}\wedge \vec{E} + \dot{\vec{B}}=0\,, \nonumber \\
\vec{\nabla}\cdot \vec{E}=& \frac{\alpha}{4\pi^{2}}\big(\vec{\nabla}\varphi\big)\cdot \vec{B}\,, \\
\vec{\nabla}\wedge \vec{B}=& \dot{\vec{E}} - \frac{\alpha}{4\pi^{2}}\lbrace \dot{\varphi}\vec{B} +\vec{\nabla}\varphi\wedge \vec{E} \rbrace\,, \nonumber
\end{align}
and
\begin{equation}\label{8.12'}
L^{-2}\,\Box\,\varphi = \frac{\alpha}{4 \pi^{2}} \vec{E}\cdot \vec{B} - \frac{\delta U(\varphi)}{\delta \varphi}.
\end{equation}
\subsection{A generalized chiral magnetic effect}
If the axion field $\varphi$ only depends on time then $\vec{\nabla}\varphi=0$, 
and, comparing the right side of \eqref{8.12} with \eqref{8.8} and reinstating $\frac{e^{2}}{\hbar}$, we find that
\begin{equation}\label{8.13}
\dot{\varphi}= \mu_{\ell}-\mu_{r}\equiv \mu_{5}\,,
\end{equation}
and \eqref{8.12} reproduces the equation describing the chiral magnetic effect; see \cite{Hehl, F-Pedrini}.

In condensed-matter theory, the equation of motion for $\dot{\varphi}\equiv \mu_5$ can take the form of a 
\textit{diffusion equation}, including a term $\tau^{-1}\mu_5$, where $\tau$ is a relaxation time associated with
the \textit{dissipation} of the asymmetry between left- and right-handed degrees of freedom. Thus
\begin{equation}\label{8.14}
\dot{\mu}_{5} + \tau^{-1}\mu_{5} - D\bigtriangleup \mu_{5} = L^{2}\frac{e^{2}}{2\pi h} \vec{E} \cdot \vec{B}\,,
\end{equation}
where $D$ is a \textit{diffusion constant}, and it is assumed here that $U(\varphi)\equiv 0$. 
Equation \eqref{8.14} implies that $\mu_{5}$ approaches
\begin{equation}\label{8.15}
\mu_{5}\simeq \frac{\tau (Le)^{2}}{2\pi h} \vec{E}\cdot \vec{B}\,,
\end{equation}
as time $t$ tends to $\infty$. This expression for $\mu_5$ can be plugged into equation \eqref{8.8} for the current
density $\vec{j}$, which then yields an expression for a conductivity tensor, $\sigma=\big(\sigma_{k\ell}\big)$, 
in the presence of an external magnetic field given by
\begin{equation}\label{8.16}
\sigma_{k\ell}=\frac{\tau(L\alpha)^{2}}{4\pi^{2}} B_{k}B_{\ell}
\end{equation}
This expression can be used in the study of transport properties of \textit{Weyl semi-metals}, to mention one example, 
which are considered in the next section.

Axion electrodynamics may have interesting applications not only in condensed-matter physics, but also 
in the theory of heavy-ion collisions, in astrophysics, and in cosmology, where it may explain the growth 
of tiny, but highly uniform cosmic magnetic fields extending over intergalactic distances, as mentioned
in Subsection 7.1; see \cite{F-Pedrini, BFR, BBFFRS} and references given there.

For some purposes, it is of interest to assume that one boundary brane, e.g., $\partial_{-}\Lambda$ (located at 
$x^4=0$), does not carry any dynamical degrees of freedom, and that $\widehat{A}|_{\partial_{-}\Lambda}=0$, 
while $\widehat{A}|_{\partial_{+}\Lambda}=:A$ is arbitrary. We then set
$$\widehat{A}_{M}(x,x^4):= \frac{x^4}{L} A(x)_{\mu},\,\, M\equiv \mu=0,1,2,3, \quad \widehat{A}_{4}(x,x^4)=: \frac{1}{L} \varphi(x).$$
The axion field $\varphi$ then transforms under an electromagnetic gauge transformations like an angle.
From the action functional \eqref{8.9} of 5D Chern-Simons electrodynamics we derive the expression for the
\textit{gauge-invariant} action functional on four-dimensional space-time $\Omega\, (=\partial_{+}\Lambda)$
\begin{align}\label{8.17}
S_{\Omega}(A,\varphi):=& \frac{1}{4\alpha} \int_{\Omega} d^{4}x\,\big[ \frac{1}{3}F_{\mu\nu}(x)F^{\mu\nu}(x) + 
\nonumber\\
+& L^{-2} \big(\partial_{\mu}\varphi(x)-A_{\mu}(x)\big) \cdot \big(\partial^{\mu}\varphi(x)-A^{\mu}(x)\big)\big] \\
+&\frac{1}{8 \pi^{2}} \int_{\Omega} \varphi \big( F\wedge F \big) + \Gamma_{\ell}(A) + \text{``irrelevant'' terms}\,.\nonumber
\end{align}
This is the effective action of an anomaly-free 4D theory of chiral fermions coupled to electromagnetism and 
to an axion-like field $\varphi$ that is \textit{not} gauge-invariant.

\section{3D Topological insulators and ``axions''}
In this last section, we study 3D topological insulators and Weyl semi-metals on a sample space-time 
$\Lambda:= \Omega\times \mathbb{R},$ with a non-empty boundary $\partial \Omega \times \mathbb{R}$. We are interested 
in the general form of the effective action describing the response of such systems to turning on an 
external electromagnetic field. Until the mid-nineties, the effective action of a 3D insulator was 
thought to be given by
\begin{equation}\label{9.1}
S_{\Lambda}(A)=\frac{1}{2}\int_{\Lambda}dt\, d^{3}x \lbrace \vec{E}\cdot \varepsilon \vec{E}- \vec{B}\cdot \mu^{-1} \vec{B} \rbrace + \text{  ``irrelevant'' terms}\,,
\end{equation}
where $\varepsilon$ is the tensor of dielectric constants and $\mu$ is the magnetic permeability tensor. The action 
defined in \eqref{9.1} is dimensionless. In the seventies, particle theorists taught us that one could add another 
dimensionless term, namely
\begin{equation}\label{9.2}
S_{\Lambda}(A) \rightarrow S_{\Lambda}^{(\theta)}(A):= S_{\Lambda}(A) + \theta\, I_{\Lambda}(A)\,,
\end{equation}
where $I_{\Lambda}$ is a \textit{``topological''} term given by
\begin{align}\label{9.3}
I_{\Lambda}(A)=&\frac{1}{4\pi^{2}}\int_{\Lambda} dt\,d^{3}x\,\vec{E}(\vec{x},t) \cdot \vec{B}(\vec{x},t)=\nonumber\\
=& \frac{1}{8\pi^{2}} \int_{\Lambda} F \wedge F \overset{Stokes}{=} \frac{1}{8\pi^{2}} \int_{\partial \Lambda} 
A\wedge dA\,.
\end{align}
In particle physics, the parameter $\theta$ is called ``vacuum angle''.
It is expected that the effective action of an insulator with broken parity and time reversal symmetries, after 
integration over all matter degrees of freedom, is given by the functional $S_{\Lambda}^{(\theta)}(A)$ introduced
in \eqref{9.2}, \eqref{9.3}. In the thermodynamic limit, $\Omega \nearrow \mathbb{R}^{3}$, 
$\text{exp}\big[iS_{\Lambda}^{(\theta)}(A)\big]$ is periodic in $\theta$ with period $2\pi$ and invariant under time reversal iff
$\theta =0, \pi$.
If $\theta=\pi$, on a space-time $\Lambda$ with a non-empty boundary $\partial \Lambda$, 
the effective action $S_{\Lambda}^{(\theta)}(A)$ contains a contribution only depending on 
$a:=A|_{\partial \Lambda}$ given by the  Chern-Simons term
\begin{equation}\label{9.4}
\Gamma_{\partial \Lambda}^{CS}(a)= \pm \frac{1}{8\pi^{2}}\int_{\partial \Lambda} a\wedge da\,,
\end{equation}
which breaks time reversal invariance. This term must be cancelled by surface degrees of freedom\footnote{I am indebted to H.-G. Zirnstein for instructive discussions of this point} on $\partial \Lambda$ exhibiting a Hall conductivity of
\begin{equation}\label{9.5}
\sigma_{H}=\mp \frac{1}{2}\cdot \frac{e^{2}}{h}\,.
\end{equation}

We have encountered the action of Eq.~\eqref{9.4} in Section 5 (see formulae \eqref{5.2}, \eqref{5.3}). 
Up to further, ``less relevant'' and time-reversal invariant terms, it is the effective action of one species 
of massless $2$-component Dirac fermions coupled to $a=A|_{\partial \Lambda}$.
\textit{Gapless quasi-particles with spin $\frac{1}{2}$} propagating along $\partial \Lambda$ could mimick such 
Dirac fermions and cancel \eqref{9.4}. It would be interesting to design 3D materials with surface degrees of 
freedom of this kind.

The vacuum angle $\theta$ might be the ground-state expectation of a dynamical field, 
$\varphi$, an \textit{``axion.''} The topological term $\theta I_{\Lambda}(A)$ would then be replaced by
\begin{equation}\label{9.6}
I_{\Lambda}(A, \varphi):= \frac{1}{8\pi^{2}}\int_{\Lambda} \varphi \big(F\wedge F\big) + S_{0}(\varphi)\,,
\end{equation}
where $S_{0}(\varphi)$ is invariant under shifts $\varphi \mapsto \varphi + 2n\pi,\,n\in \mathbb{Z}$.
We then enter the realm of \textit{axion-electrodynamics}, as reviewed in Section 8. Recalling the field equations 
\eqref{8.12}, we find \textit{Halperin's ``3D quantum Hall effect''}:
From \eqref{8.12} we infer a formula for the current $\vec{j}$ generated in an electromagnetic field, namely
\begin{equation}\label{9.7}
\vec{j}= -\frac{e^{2}}{4\pi h}\left( \dot{\varphi} \cdot \vec{B} + \vec{\nabla} \varphi\times \vec{E} \right)
\end{equation}
Consider a 3D \textit{spatially periodic} system of matter (with crystal lattice $\frak{L}$) exhibiting an effective 
axion $\varphi$. We suppose that $\varphi$ is time-independent, i.e., $\mu_5 =0$. Taking into account the 
periodicity of $\text{exp}\big(i I_{\Lambda}(A, \varphi)\big)$ under shifts, 
$\varphi \mapsto \varphi + 2n\pi, n \in \mathbb{Z}$, invariance under lattice translations 
implies that
\begin{equation}\label{9.7}
\varphi(\vec{x})=2\pi \,\big(\vec{K}\cdot \vec{x}\big) + \phi(\vec{x})\,,
\end{equation}
where $\vec{K}$ belongs to the \textit{dual lattice} $\frak{L}^{*}$, and $\phi$ is invariant under lattice translations. Neglecting 
$\phi$, we find that $ \vec{\nabla}\varphi= 2\pi \vec{K}$\, is ``quantized,'' as $\vec{K}$ belongs to the \textit{dual}
of the crystal lattice $\frak{L}$. With \eqref{9.7} this implies
that 
$$\vec{j}= \frac{e^{2}}{2h}\vec{K}\times \vec{E}\,, \qquad \vec{K}\in \frak{L}^{*}\,.$$
This is Halperin's 3D \textit{(quantum) Hall effect} \cite{Halperin-2}.\footnote{I am indebted to Greg Moore for
a very instructive discussion of this effect.} \\

Next, I describe ``axionic'' effects in topological insulators with an effective action given by 
(see \eqref{9.1}, \eqref{9.2} and \eqref{9.6}) 
\begin{equation}\label{9.9}
S_{\Lambda}(A,\varphi)=S_{\Lambda}(A)+ \frac{1}{8\pi^{2}}\int_{\Lambda} \varphi F\wedge F + S_{0}(\varphi)\,,
\end{equation}
where $S_{0}(\varphi)$ is now assumed to be invariant under shifts $\varphi \mapsto \varphi + n\pi,\, n\in \mathbb{Z}$. 
It is compatible with time-reversal invariance that $S_{0}(\varphi)$ has (well developed) minima at $\varphi=n\pi$. 
Then the material described by \eqref{9.9} is \textit{not} an ordinary insulator. It may exhibit a 
\textit{Mott transition}: At a positive temperature, the bulk of such a material contains \textit{domain walls} 
across which the value of the axion field $\varphi$ jumps by an integer multiple of $\pi$. 
Recalling the insight described after \eqref{9.4} and \eqref{9.5}, we predict that such domain walls 
may carry gapless two-component Dirac-type fermions. At sufficiently high temperatures, some of
these domain walls can be expected to become macroscopic, and this would then give rise to a 
\textit{non-vanishing conductivity}; see \cite{F-Werner}.

A curious effect related to the one described in Subsection 7.1 (see also \cite{F-Pedrini}) that could possibly 
be observed in materials with an effective axion has been pointed in \cite{O-O}: A time-independent 
external \textit{electric field} $\vec{E}$ applied to an axionic magnetic material is \textit{screened} once its 
strength exceeds a certain critical value, the excess energy being fed into the growth of a \textit{magnetic field}.

\subsection{Weyl semi-metals}

It may be argued that dynamical axions could emerge as effective degrees of freedom in
\begin{itemize}
\item{certain \textit{3D topological insulators} with anti-ferromagnetic short-range order, (magnetic 
fluctuations playing the role of a dynamical axion);\footnote{a conjecture proposed by S.-C. Zhang 
and coworkers \cite{Zhang}, inspired by the work in \cite{F-Pedrini}} and in}
\item{crystalline \textit{3D Weyl semi-metals},}
\end{itemize}
i.e., in systems with two energy bands exhibiting two (or, more generally, an even number\footnote{This folllows 
from the celebrated Nielsen-Ninomiya theorem; see \cite{Friedan} and references given there} of) double-cones in quasi-momentum (frequency) space 
corresponding to \textit{chiral} quasi-particle states, assuming that the Fermi energy is close to the apices 
of those double-cones; see  \cite{Herring, Hasan}. At low frequencies, namely near the apices of those 
double-cones, the quasi-particle states of such systems are left- or right-handed Weyl fermions, respectively. 
In such systems, the time derivative, $\dot{\varphi}\equiv \mu_5$, of the ``axion field'' $\varphi$ 
really has the meaning of a time-dependent \textit{difference of chemical potentials} of left-handed 
and right-handed quasi-particles. It satisfies an equation of motion of the kind described in \eqref{8.14}; i.e.,
\begin{equation}\label{9.8}
\dot{\mu}_{5} + \tau^{-1}\mu_{5} - D\bigtriangleup \mu_{5} = L^{2}\frac{e^{2}}{2\pi h} \vec{E} \cdot \vec{B}\,.
\end{equation}
A non-vanishing initial value of the chemical potential $\mu_5$ may be triggered by strain applied to the system, 
leading to a left-right asymmetric population of the Fermi sea. Due to \textit{``inter-valley'' 
scattering processes}, a non-vanishing $\mu_5$ will then relax towards $0$, with a relaxation time corresponding 
to the parameter $\tau$ in Equation \eqref{9.8}. However, when applying a small external electric field 
$\vec{E}$ and a magnetic induction $\vec{B}$ to the system, with the property that 
$\vec{E}\cdot \vec{B}\not= 0$, one finds from \eqref{9.8} that the potential $\mu_5$ 
relaxes towards $\mu_5\simeq \frac{\tau (Le)^{2}}{2\pi h} \vec{E}\cdot \vec{B}$. 
Thus, the conductivity tensor, $\sigma=\big(\sigma_{k\ell}\big)_{k,\ell=1,2,3}$, is given by
$$\sigma_{k\ell}= \sigma^{(0)}_{k\ell} + \frac{\tau(L\alpha)^{2}}{4\pi^{2}} B_{k}B_{\ell}\,, $$
where the first term on the right side is the standard Ohmic conductivity (due to phonon- and impurity scattering), 
and the second term is a manifestation of the \textit{chiral magnetic effect}.

\section{Conclusions and acknowledgements}
I conclude this paper with a few general remarks, including ones about what is missing in this review and some
more personal ones.
\begin{enumerate}
\item{I hope that this review shows that the physics of systems of condensed matter in two and three 
dimensions, and in particular of (topological) insulators, is very rich. It has a lot of potential to find 
promising technological applications. Mathematical techniques ranging from abstract algebra 
over the topology of fiber bundles all the way to hard analysis have interesting and often surprising 
applications in the study of these fascinating states of matter. Electron liquids, magnetic 
materials, Bose gases, etc.~offer many challenges to experimentalists and theorists. General 
principles and interesting features of quantum theory, such as braid statistics \cite{braids}, fractional 
spin and fractional electric charges (see, e.g., \cite{F-Marchetti} and references given there), anomalies 
and their cancellation, current algebra and holography have appearances in the arena of condensed matter 
physics. Exotic species of quasi-particles, in particular, quasi-particles with braid statistics, two-component 
Dirac-like fermions and Majorana fermions are encountered in the physics of various two- and 
three-dimensional systems of condensed matter explored in recent years.}
\item{My main goal in this paper has been to review some of the results in condensed-matter physics
obtained by my collaborators and myself that are \textit{not} limited to non-interacting electron 
gases, but apply to correlated systems exhibiting interaction effects. I hope this review shows 
that, apparently, concepts and methods from quantum field theory pioneered in particle 
physics can be used to one's advantge to study general features of \textit{interacting} systems 
in condensed-matter physics. In particular, these methods are very useful to characterize and classify 
plenty of examples of \textit{``topological states of matter,''} in particular ones that do not have any local 
order parameters so that Landau theory is not applicable. (Likely the best examples are 2D incompressible 
electron liquids exhibiting the fractional quantum Hall effect; see \cite{Prange, F-Thiran, Les Houches 94} and 
references given there.) In this paper this is illustrated by showing how concepts from \textit{quantum gauge theory}, 
such as effective actions, power counting, gauge invariance, the chiral anomaly and some of its consequences, 
such as the chiral magnetic effect, $\theta$ - ground-states, axion electrodynamics, quasi-particles with
braid statistics, etc.~can be used to come up with non-trivial and often quite surprising insights into 
properties of states of systems of condensed matter that have attracted much interest in recent years. 
The general formalism reviewed in this paper can thus rightly be called \textit{Gauge Theory of States/Phases of Matter}.
The ambitious goals of our efforts may explain why mathematical rigor has not been the
main concern in our efforts -- the mathematical complexity of quantum-mechanical many-body theory 
of interacting/correlated systems is well known to be awesome. A certain lack of mathematical rigor inherent in 
many of our results may be a reason why they have not become widely appreciated among mathematical 
physicists. And the inspiration we have drawn from particle physics and relativistic quantum field theory 
may explain why our results have been largely ignored among condensed-matter theorists.\footnote{To read, 
to mention but one example (see \cite{Landsteiner}), that it is \textit{only during the past decade} 
that the chiral anomaly has been used in the study of transport phenomena in condensed matter 
physics makes me wonder whether communication channels within the scientific community are 
congested (to put it politely). Of course, there are also examples of this congestion in the mathematical physics
community. -- In all modesty I claim some credit for my collaborators and myself for having developed 
a useful formalism, emphasizing connections to particle physics and quantum field theory including 
gauge theory and the chiral anomaly, for the study of ``topological states of matter,'' at least thirty 
years ago; see \mbox{\cite{F-Marchetti} - \cite{Faddeev}} and \cite{braids}. } -- 
It is hoped that this paper may help to change the situation.
}
\item{
What is clearly and most regrettably missing in this review is an account of the \textit{bare-hands analysis} 
of spectral properties of many-body Hamiltonians describing ``topological states of matter'' at energies 
quite close to the ground-state energy and to derive properties of quasi-particles and of transport, using 
a variety of powerful analytical tools, such as multi-scale analysis\footnote{a technique that owes some of
its early momentum to a paper \cite{F-Spencer} containing the first mathematically rigorous proof of existence 
of the Berezinskii-Kosterlitz-Thouless transition} and renormalization group methods. 
Colleagues who have devoted serious efforts extending over many years towards reaching results in this direction 
are  T.~Balaban, F.~Dyson, J. Feldman, G. Gallavotti, A. Giuliani, H. Kn\"{o}rrer, E.~H.~Lieb, J.~Magnen, 
V. Mastropietro, W.~Pedra, A.~Pizzo, M. Porta, V.~Rivasseau, M.~Salmhofer, B.~Schlein, R.~Seiringer, 
E. Trubowitz, J.~Yngvason, and others. 
References to early work by these colleagues can be found in my Les Houches lectures \cite{Les Houches 94-2} 
of 1994; among a variety of recent contributions towards understanding analytical aspects of the quantum Hall effect are
the papers quoted in \cite{Mastropietro}. Further references to many important results in mathematical
quantum many-body theory, due to colleagues mentioned in this paper and numerous other researchers, 
can easily be found on `arXiv' and `google scholar.' I strongly recommend their papers to the attention of 
the reader. The purpose of this paper and of the bulk of the references given in the text is, however, to 
draw the readers' attention to work of my collaborators and myself, rather than to provide a comprehensive 
survey of the field featured in it. -- Obviously plenty of fascinating open problems remain to be tackled.}
\end{enumerate}

\textit{Acknowledgements.} I am immensely grateful to my many collaborators who have joined me on our 
sometimes arduous journey through some provinces of condensed-matter physics, including the following
colleagues: A.~Alekseev, S.~Bieri, A.~Boyarsky, V.~Cheianov, T.~Chen, R.~G\"otschmann, \mbox{G.-M.} Graf, 
T.~Kerler, I.~Levkivskyi, P.-A.~Marchetti, B.~Pedrini, A.~Pizzo, O.~Ruchayskiy, Chr. Schweigert, T.~Spencer,
U.~M.~Studer, E.~Sukhorukov, E.~Thiran, J.~Walcher, Ph.~Werner, and A.~Zee. I am very grateful to the 
late R.~Morf for having introduced me to the miracles of the quantum Hall effect, for many discussions 
and for his criticism, to P.~Wiegmann for very helpful discussions and encouragement, and, 
last but not least, to M.~Zirnbauer for his kind interest, extending over many years, in my efforts, and 
for having invited me to a workshop at the Schr\"odinger Institute in Vienna and for a series of lectures at 
Bad Honnef, on which this review is based. I have profited from what colleagues at ETH Zurich and 
elsewhere -- too many to be listed here -- have taught me in numerous discussions.\\

AIPP-Data Availability Statement: Data sharing is not applicable to this article as no new data 
were created or analyzed in this study.

\end{document}